\documentclass[onecolumn,twocolumn]{IEEEtran}
\usepackage[latin9]{inputenc}
\usepackage{color}
\definecolor{page_backgroundcolor}{rgb}{1, 1, 1}
\pagecolor{page_backgroundcolor}
\definecolor{document_fontcolor}{rgb}{0, 0.335938, 1}
\color{document_fontcolor}
\usepackage{float}
\usepackage{bm}
\usepackage{amsmath}
\usepackage{amsthm}
\usepackage{amssymb}
\usepackage{graphicx}

\makeatletter

\floatstyle{ruled}
\newfloat{algorithm}{tbp}{loa}
\providecommand{\algorithmname}{Algorithm}
\floatname{algorithm}{\protect\algorithmname}

\theoremstyle{plain}
\newtheorem{thm}{\protect\theoremname}
\theoremstyle{plain}
\newtheorem{prop}[thm]{\protect\propositionname}
\theoremstyle{plain}
\newtheorem{lem}[thm]{\protect\lemmaname}

\IEEEoverridecommandlockouts
\usepackage{cite}
\usepackage{amsfonts}\usepackage{algorithmic}
\usepackage{textcomp}
\usepackage{xcolor}
\def\BibTeX{{\rm B\kern-.05em{\sc i\kern-.025em b}\kern-.08em
    T\kern-.1667em\lower.7ex\hbox{E}\kern-.125emX}}
\usepackage[acronym]{glossaries}
\newcommand{\newac}{\newacronym}
\newcommand{\ac}{\gls}

\makeglossaries
\newac{isac}{ISAC}{sensing and communication}
\newac{an}{AN}{artificial noise}
\newac{csi}{CSI}{channel state information}
\newac{csit}{CSIT}{channel state information at transmitter}
\newac{bme}{BME}{beampattern matching error}
\newac{ula}{ULA}{uniform linear array}
\newac{cu}{CU}{communication user}
\newac{sinr}{SINR}{signal-to-interference-plus-noise ratio}
\newac{sdr}{SDR}{semi-definite relaxation}
\newac{1d}{1D}{one-dimensional}
\newac{zf}{ZF}{zero-forcing}
\newac{aod}{AoD}{angle of departure}
\newac{los}{LoS}{line-of-sight} 
\newac{nlos}{NLoS}{non-line-of-sight} 
\newac{bs}{BS}{base station}
\newac{bss}{BSs}{base stations}
\newac{cus}{CUs}{communication users}
\newac{snr}{SNR}{signal-to-noise ratio}
\newac{evd}{EVD}{eigenvalue decomposition}
\newac{qsdp}{QSDP}{quadratic semidefinite programing}
\newac{dof}{DoF}{degrees-of-freedom}
\newac{soc}{SOC}{second order cone}
\newac{cscg}{CSCG}{circularly symmetric complex Gaussian}
\newac{crb}{CRB}{Cram{\'e}r-Rao bound}

\providecommand{\propositionname}{Proposition}
\providecommand{\theoremname}{Theorem}

\makeatother

\providecommand{\lemmaname}{Lemma}
\providecommand{\propositionname}{Proposition}
\providecommand{\theoremname}{Theorem}

\begin{document}
\title{Robust Secure Communications in Near-Field ISCAP Systems with Extremely
Large-Scale Antenna Array}
\author{\IEEEauthorblockN{Zixiang Ren, Siyao Zhang, Ling Qiu,{\large{} Derrick Wing Kwan Ng, and
}Jie Xu,}\vspace{-0.6cm}\thanks{Part of this paper has been presented at IEEE International Symposium
on Wireless Communication Systems (ISWCS) 2024 \cite{ren2024secure}.}\thanks{Z. Ren is with the Key Laboratory of Wireless-Optical Communications,
Chinese Academy of Sciences, School of Information Science and Technology,
University of Science and Technology of China, Hefei 230027, China,
and also with  the Shenzhen Future Network of Intelligence Institute
(FNii-Shenzhen), and the Guangdong Provincial Key Laboratory of Future
Networks of Intelligence, The Chinese University of Hong Kong, Shenzhen,
Guangdong 518172, China (e-mail: rzx66@mail.ustc.edu.cn).}\thanks{S. Zhang is with the FNii-Shenzhen and the Guangdong Provincial Key
Laboratory of Future Networks of Intelligence, The Chinese University
of Hong Kong, Shenzhen, Guangdong 518172, China (e-mail: zsy@mails.swust.edu.cn).}\thanks{L. Qiu is with the Key Laboratory of Wireless-Optical Communications,
Chinese Academy of Sciences, School of Information Science and Technology,
University of Science and Technology of China, Hefei 230027, China.
(e-mail: lqiu@ustc.edu.cn).}\thanks{D. W. K. Ng is with the School of Electrical Engineering and Telecommunications,
University of New South Wales, Sydney, NSW 2052, Australia (e-mail:
w.k.ng@unsw.edu.au).}\thanks{J. Xu is with the School of Science and Engineering (SSE), the FNii-Shenzhen,
and the Guangdong Provincial Key Laboratory of Future Networks of
Intelligence, The Chinese University of Hong Kong, Shenzhen, Guangdong
518172, China (e-mail: xujie@cuhk.edu.cn).}\thanks{J. Xu and L. Qiu are the corresponding authors.}}
\maketitle
\begin{abstract}
This paper investigates robust secure communications in a near-field
integrated sensing, communication, and powering (ISCAP) system, in
which the base station (BS) is equipped with an extremely large-scale
antenna array (ELAA). In this system, the BS transmits confidential
messages to a single legitimate communication user (CU), simultaneously
providing wireless power transfer to multiple energy receivers (ERs)
and performing point target sensing. We consider a scenario in which
both the ERs and the sensing target may act as potential eavesdroppers
attempting to intercept the confidential messages. To safeguard secure
communication, the BS employs a joint beamforming design by transmitting
information beams combined with dedicated triple-purpose beams serving
as energy and sensing signals, as well as artificial noise (AN) for
effectively jamming potential eavesdroppers. It is assumed that only
coarse location information of the ERs and sensing targets or eavesdroppers
is available at the BS, leading to imperfect channel state information
(CSI). Under this setup, we formulate a robust beamforming optimization
problem with the objective of maximizing the secrecy rate for the
CU, while ensuring worst-case performance requirements on both target
sensing and wireless energy harvesting at the ERs. To address the
non-convex robust joint beamforming problem and facilitate the deployment
of a low-complexity algorithm, we employ the S-procedure alongside
an eavesdropping CSI error-bound determination method to acquire a
high-quality solution. Numerical results demonstrate that the proposed
optimization framework can effectively maximizes secrecy rates by
leveraging near-field beamforming's spatial focusing to suppress eavesdropper
leakage, while its robust design guarantees worst-case sensing and
wireless power transfer performance, concurrently demonstrating synergistic
gains for precise localization. 
\end{abstract}


\section{Introduction}

Integrated sensing, communication, and powering (ISCAP) has been recognized
as a cornerstone technology to facilitate multi-functional operations
in future sixth-generation (6G) wireless networks, in which wireless
signals concurrently support the triple roles of sensing, communication,
and wireless power transfer (WPT) \cite{chen2024integrated}. By seamlessly
integrating these functions into a unified system, ISCAP offers a
comprehensive solution capable of addressing the increasing complex
demands of emerging network applications. In practice, the benefits
of ISCAP are manifold \cite{chen2024isac}. First, it enhances spectrum
utilization efficiency by allowing simultaneous operations within
the same frequency bands, thus reducing the need for separate spectral
resources dedicated to each function individually. Second, by integrating
WPT, ISCAP supports sustainable operation of energy-constrained devices,
which is particularly crucial for large-scale Internet-of-Things (IoT)
deployments. Third, the inclusion of sensing capabilities within the
network infrastructure allows real-time environmental monitoring and
improved situational awareness, which are essential for various applications
such as autonomous systems and smart urban environments. Given these
substantial benefits, ISCAP is expected to operate under several distinct
modes, such as simultaneous multi-functional transmission \cite{chen2024isac},
sensing-assisted wireless information and power transfer (WIPT) \cite{zhang2024training,xu2024sensing},
networked ISACP, and wirelessly powered integrated sensing and communication
(ISAC) \cite{liu2022integrated,liu2020joint}. Among these, simultaneous
multi-functional transmission enables the concurrent execution of
sensing, communication, and power transfer by leveraging shared radio
frequency (RF) resources. This mode is particularly appealing to satisfy
the intricate demands for both sensing and communication in future
sustainable 6G networks, by jointly optimizing resource utilization
efficiency, service quality, and energy efficiency, which is thus
the primary focus of this work.

The simultaneous multi-functional transmission introduces both challenges
and opportunities regarding communication security. Specifically,
the integration of sensing and WPT exploiting communication signals
inherently increases security vulnerabilities, as energy receivers
(ERs) or sensing targets could potentially exploit the overheard beamformed
signals to eavesdrop on confidential information \cite{zhu2024enabling,ren2023robust}.
To mitigate these risks, physical layer security methods have been
developed to safeguard confidential message transmissions by exploiting
intrinsic properties of wireless channels \cite{shiu2011physical}.
In practice, the secrecy rate serves as a fundamental metric for evaluating
communication security, which is defined as the maximum achievable
secure communication rate that prevents interception of eavesdroppers
\cite{chen2016survey,mukherjee2014principles}. To fully unlock the
potential of physical layer security, various advanced techniques
such as secure beamforming and artificial noise (AN) injection have
been proposed (e.g., \cite{goel2008guaranteeing}). Indeed, secure
beamforming exploits spatial beamforming to adaptively steer information
beams to intended communication users (CUs) while simultaneously nullifying
the reception of potential eavesdroppers \cite{chen2016survey}. In
the context of ISCAP systems, secure beamforming can also be applied
to facilitate achieving different system objectives, such as optimizing
energy delivery to legitimate ERs or enhancing sensing capabilities,
thereby maintaining a multi-functional balance \cite{liu2014secrecy,su2020secure}.
To further strengthen security and complementary secure beamforming,
AN injection serves as an efficient physical layer security technique
\cite{liu2014secrecy,liu2018exploiting}. In this approach, AN is
intentionally injected into the wireless channel to degrade reception
quality at eavesdroppers, while minimally affecting legitimate receivers
\cite{shiu2011physical}. Remarkably, in ISCAP systems, AN can also
be strategically reused to provide sensing and powering functionalities
\cite{su2020secure,chen2016secrecy,ng2014robust}. Consequently, the
synergistic integration of secure beamforming and AN in ISCAP systems
not only enhances physical-layer-security but also optimizes overall
performance across these multi-functional operations. 

On the other hand, extremely large-scale antenna array (ELAA) has
emerged as another disruptive technique for 6G networks, offering
substantial benefits for sensing, communication, and powering \cite{lu2023near}.
In fact, a fundamental paradigm shift introduced by ELAA is the transition
from traditional plane-wave propagation models to spherical-wave propagation
in the near-field region \cite{cui2022channel}. This transition arises
due to the significantly increased aperture sizes associated with
ELAA, which extends the near-field boundary to practical distances
for typical deployment scenarios \cite{cui2022near}. Unlike plane-wave
propagation models, which assume parallel wavefronts and simplify
channel characterization, spherical-wave models explicitly account
for wavefront curvature, enabling precise beamfocusing across both
angular and distance dimensions \cite{zhang2022beam}. Consequently,
ELAA systems can effectively enhance communication through high-gain
beamforming, enable precise sensing via ultra-narrow beams, and improve
WPT efficiency through accurately focusing energy onto targeted ERs
\cite{wang2023near,garnica2013wireless,qu2023near,lu2023beamforming}.
Furthermore, the enhanced resolution in the distance domain provided
by near-field communication techniques can be exploited to significantly
strengthen physical layer security \cite{zhang2024near}. In particular,
by leveraging the precise distance estimation capabilities enabled
by spherical wavefront propagation, the system can better localize
and mitigate potential eavesdropping, thereby improving the robustness
of secure transmission \cite{zhang2024physical}. 

While ELAA architectures enable transformative near-field ISCAP, precise
channel state information (CSI) acquisition remains critical to realize
their theoretical beamforming gains. Accurate CSI estimation enables
the optimization of triple-functional beam patterns for joint communication
spectral efficiency, sensing resolution, and wireless power transfer.
For legitimate users, orthogonal pilots with codebook-based limited
feedback serve as the baseline CSI acquisition method. In contrast,
estimating CSI for ERs and potential eavesdroppers presents significant
challenges due to their inherent passive nature, often requiring advanced
blind estimation techniques or proactive security mechanisms to mitigate
their hidden presence \cite{chen2016secrecy,liu2018exploiting}. Consequently,
such inherent uncertainties leads to inevitable CSI errors for these
passive nodes, which directly affects the effectiveness of security-focused
beamforming design. As a result, it is important to incorporate robust
design principles into secure ISCAP by considering either worst-case
scenarios or probabilistic models for CSI uncertainties \cite{ng2014robust,ren2023robust}.
This robust design approach has been widely adopted in various research
efforts of secure WIPT or ISAC systems. For example, the authors in
\cite{ng2014robust} investigated robust secure beamforming for multi-user
multi-input-single-output (MISO) WIPT systems to maximize the worst-case
secrecy rate by considering imperfect CSI of the potential eavesdroppers.
Their approach achieves the maximum worst-case secrecy rate and guarantees
energy harvesting requirements in WIPT systems by explicitly accounting
for bounded eavesdroppers' CSI errors. Additionally, for secure ISAC
design, the authors in \cite{ren2023robust} investigated robust secure
beamforming in a single-user multi-target ISAC system. They maximized
the sensing beampattern matching performance while ensuring the worst-case
secrecy rate under two practical eavesdropper CSI error scenarios
with bounded and Gaussian errors, respectively. 

Existing robust beamforming paradigms remain constrained to WIPT/ISAC
frameworks \cite{ng2014robust,ren2023robust}, inadequately addressing
near-field ISCAP's unique requirements. Conventional bounded/Gaussian
error models fail to capture spherical wavefront effects inherent
to extremely large apertures. More critically, the physical basis
for parameterizing uncertainty sets, particularly the calibration
of error bounds against near-field \cite{xu2022robust,HuangLeeJ12}
remains unestablished. The fundamental challenge lies in simultaneously
optimizing secure spectral efficiency, sensing precision, and power
transfer under location-dependent CSI with parameters, requiring novel
robustness frameworks against multi-dimensional uncertainties of eavesdroppers.
This tri-objective optimization objective constitutes a critical knowledge
gap, motivating our investigation into secure ISCAP systems.

This paper investigates robust secure communications in a near-field
ISCAP system where an ELAA is deployed at the base station (BS). We
consider a scenario of simultaneous multi-functional transmission
of ISCAP, in which the BS simultaneously transmits confidential messages
to a legitimate CU, transfer energy to multiple ERs, and performs
sensing of a point target. In this setting, the ERs and the sensing
target potentially serve as potential eavesdroppers, and the BS has
access only to coarse location information of these entities. The
primary contributions of this paper are summarized as follows.
\begin{itemize}
\item We first develop a robust joint beamforming framework to maximize
the secrecy rate of the CU while guaranteeing the worst-case performance
in terms of sensing accuracy and energy harvesting. Specifically,
dedicated auxiliary beams are carefully designed to concurrently serve
a triple role of as energy signals, sensing signals, and AN, effectively
jamming potential eavesdroppers and enhancing security. These beams
exploit the unique spatial resolution properties of near-field ELAA
to achieve robust performance even under the presence of uncertain
conditions. 
\item We formulate a robust joint beamforming optimization problem to maximize
the secrecy rate while guaranteeing the worst-case performance for
both energy harvesting and sensing accuracy (via the \ac{crb} constraints).
Our problem formulation explicitly accounts for practical location
estimation errors that induce channel uncertainty. The resulting problem
presents significant analytical and computational challenges due to
the implicit nature of the location-induced channel error model, the
inherent non-convexity of both the secrecy rate objective function,
and the coupled security/sensing constraints. 
\item To address this non-convex optimization problem, we first employ semidefinite
relaxation (SDR) and the Charnes-Cooper transformation to reformulate
the secrecy rate maximization into a tractable form. Subsequently,
a novel channel error decomposition method is developed, leveraging
Taylor series approximations, enabling us to bound location-induced
CSI uncertainties into geometric (or line-of-sight (LoS)) and non-line-of-sight
(NLoS) components. Then, the S-procedure is applied to transform worst-case
constraints into linear matrix inequalities (LMIs), while a one-dimensional
(1D) search combined with fractional programming ensures convergence.
Our solution preserves SDR tightness, rigorously guaranteeing the
performance under all imposed constraints for the reformulated problem.
\item Finally, numerical results clearly demonstrate the efficacy of our
proposed method in capitalizing the unique spatial near-field characteristics
of ELAA to significantly enhance secure communication performance
compared to three benchmarks. Specifically, the proposed optimization
framework effectively maximizes the secrecy rate by exploiting the
precise spatial focusing capabilities offered by near-field beam focusing,
thereby substantially reducing signal leakage to potential eavesdroppers.
Meanwhile, the robust design ensures worst-case performance guarantees
for both sensing and energy transfer functionalities. The results
also highlight the synergistic benefits of near-field beamforming
for accurate localization.
\end{itemize}

\textit{Notations}: Vectors and matrices are denoted by bold lower-
and upper-case letters, respectively. $\mathbb{C}^{N\times M}$ denotes
the space of $N\times M$ complex matrices. $\boldsymbol{I}$ and
$\boldsymbol{0}$ represent an identity matrix and an all-zero matrix
with appropriate dimensions, respectively. For a square matrix $\boldsymbol{A}$,
$\textrm{tr}(\boldsymbol{A})$ denotes its trace and $\boldsymbol{A}\succeq\boldsymbol{0}$
means that $\boldsymbol{A}$ is positive semi-definite. For a complex
arbitrary-size matrix $\boldsymbol{B}$, $\boldsymbol{B}[m,n]$, $\textrm{rank}(\boldsymbol{B})$,
$\boldsymbol{B}^{T}$, $\boldsymbol{B}^{H}$, and $\boldsymbol{B}^{c}$
denote its $(m,n)$-th element, rank, transpose, conjugate transpose,
and complex conjugate, respectively, and $\mathrm{vec}(\boldsymbol{B})$
denotes the vectorization of $\boldsymbol{B}$. For a vector $\boldsymbol{a}$,
$\boldsymbol{a}[i]$ denotes its $i$-th element. $\mathbb{E}(\cdot)$
denotes the statistical expectation. $\|\cdot\|$ denotes the Euclidean
norm of a vector. $|\cdot|$, $\mathrm{Re}(\cdot)$, and $\mathrm{Im}(\cdot)$
denote the absolute value, the real component, and the imaginary component
of a complex entry. $\mathcal{CN}(\boldsymbol{x},\boldsymbol{Y})$
denotes a \ac{cscg} random vector with mean vector $\boldsymbol{x}$
and covariance matrix $\boldsymbol{Y}$. $\boldsymbol{A}\odot\boldsymbol{B}$
represent the Hadamard product of two matrices $\boldsymbol{A}$ and
$\boldsymbol{B}$. $\frac{\partial}{\partial(\cdot)}$ denotes the
partial derivative operator and $\nabla(\cdot)$ denotes the gradient
operator. $j=\sqrt{-1}$. $(x)^{+}=\max(x,0)$.

\section{System Model and Problem Formulation}

\begin{figure}
\vspace{-0.4cm}\includegraphics[scale=0.15]{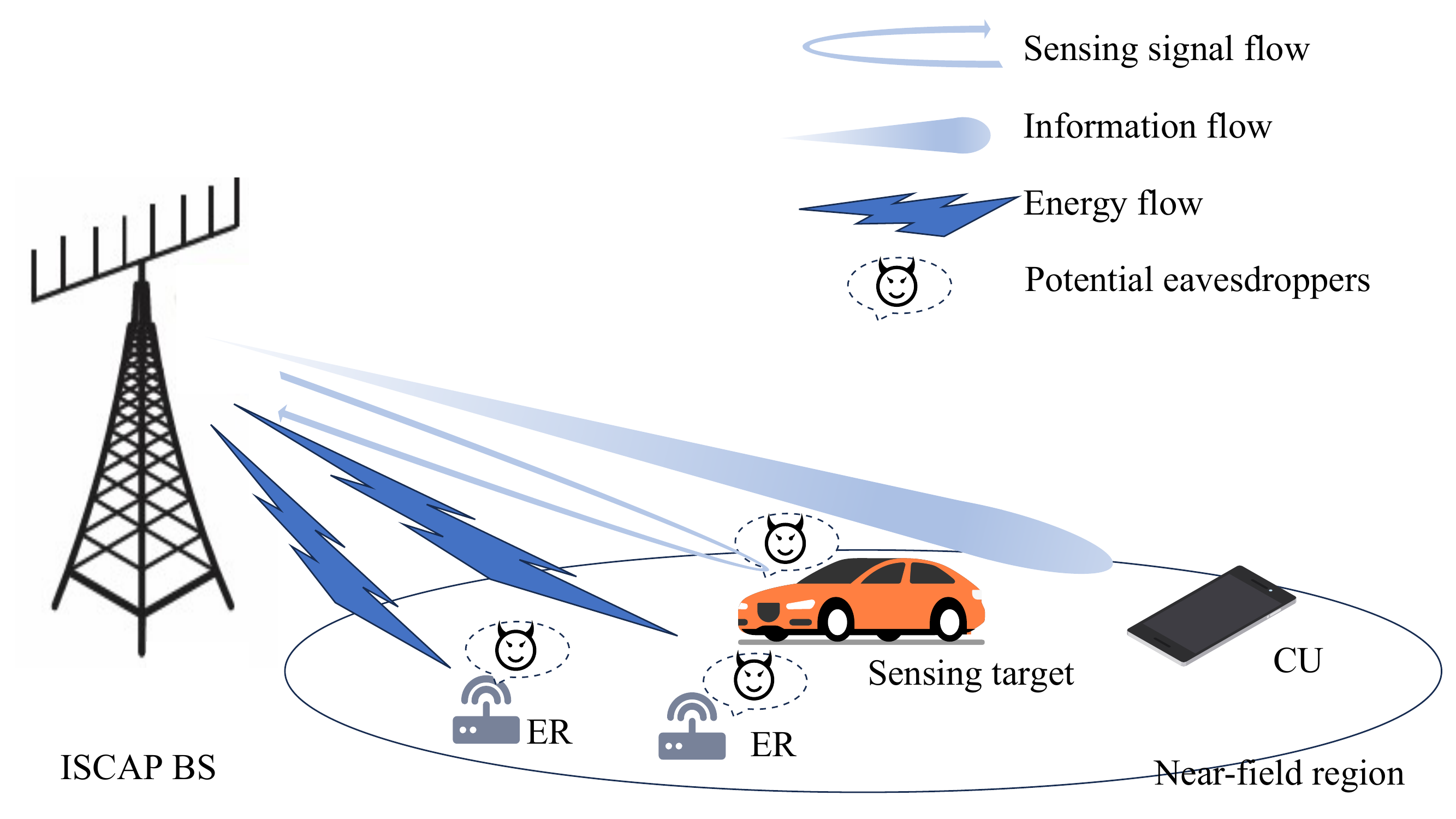}\centering\caption{Illustration of the considered ISCAP system.}
\vspace{-0.4cm}
\end{figure}

This paper considers a secure ISCAP system as shown in Fig. 1, which
comprises a multi-functional BS, one sensing target, $K$ single-antenna
ERs, and a single-antenna CU. We assume that the BS is equipped with
a \ac{ula} of $N$ transmit antennas with antenna spacing $d$,
and the CU employs a single antenna. All terminals (the CU and the
ERs) and the sensing target are located in the near-field region of
the BS, i.e., their distances from the BS are less than the Rayleigh
distance $2D^{2}/\lambda$, where $D$ denotes the array aperture
and $\lambda$ represents the wavelength of the adopted signal carrier
frequency \cite{cui2022near}. The BS concurrently performs three
critical functions: 1) confidential information delivery to the CU,
2) wireless power transfer to the ERs, and 3) target localization
via sensing. Notably, both ERs and the sensing target are considered
as potential eavesdroppers attempting to intercept confidential messages.
Let $\mathcal{K}_{\mathrm{ER}}\overset{\triangle}{=}\{1,2,\dots,K\}$
denote the set of all $K$ ERs and $\mathcal{K}_{\mathrm{EAV}}=\mathcal{K}_{\mathrm{ER}}\cup\{K+1\}$
denote the set of potential eavesdroppers, in which $k=K+1$ represents
the sensing target.

We focus on the secure ISCAP over a transmission block of $T$ symbols.
At each symbol period $t\in\{1,\dots,T\}$, the BS transmits a confidential
message, $s_{0}(t)\in\mathbb{C}$, to the CU through a dedicated beamforming
vector $\boldsymbol{w}_{0}\in\mathbb{C}^{N\times1}$, where $s_{0}(t)$
is a CSCG random variable with zero mean and unit variance, i.e.,
$s_{0}(t)\sim\mathcal{CN}(0,1)$. To facilitate secure ISCAP, the
BS simultaneously emits a dedicated signal vector $\boldsymbol{s}_{1}(t)\in\mathbb{C}^{N\times1}$,
serving three purposes of energy transmission, target sensing, and
AN generation. We assume that $\boldsymbol{s}_{1}(t)$ is independent
from $s_{0}(t)$, $\forall t\in\{1,\dots,T\}$, and the former follows
a CSCG distribution with zero mean and a general covariance matrix
$\boldsymbol{R}_{1}=\mathbb{E}(\boldsymbol{s}_{1}(t)\boldsymbol{s}_{1}^{H}(t))\succeq\boldsymbol{0},\textrm{i.e}.,$
$\boldsymbol{s}_{1}(t)\sim\mathcal{CN}(\boldsymbol{0},\boldsymbol{R}_{1})$.
The optimizable covariance matrix $\boldsymbol{R}_{1}$ is assumed
to be of an arbitrary rank with $0\le\mathrm{rank}(\boldsymbol{R}_{1})\le N$,
enabling flexible beamspace resources management through eigenvalue
decomposition. As a result, the transmitted signal by the BS is expressed
as
\begin{equation}
\boldsymbol{x}(t)=\boldsymbol{w}_{0}s_{0}(t)+\boldsymbol{s}_{1}(t).
\end{equation}
Consequently, the transmit covariance matrix of $\boldsymbol{x}(t)$
is
\begin{equation}
\boldsymbol{R}=\mathbb{E}(\boldsymbol{x}(t)\boldsymbol{x}^{H}(t))=\boldsymbol{R}_{0}+\boldsymbol{R}_{1},\label{eq:covariance}
\end{equation}
where $\boldsymbol{R}_{0}=\boldsymbol{w}_{0}\boldsymbol{w}_{0}^{H}$
with $\boldsymbol{R}_{0}\succeq\boldsymbol{0}$ and $\textrm{rank}(\boldsymbol{R}_{0})\leq1$.
We consider that the BS operates under a transmit power budget $P$,
resulting in 
\begin{equation}
\mathrm{Tr}(\boldsymbol{R}_{0}+\boldsymbol{R}_{1})\leq P.\label{eq:power constraint}
\end{equation}
Then, we consider the geometric configuration of the ELAA. Without
loss of generality, we assume that the ULA is oriented along the plane
of x-axis, centered at the origin. Accordingly, the Cartesian coordinate
vector of its $n$-th antenna element is denoted as $\boldsymbol{u}_{n}=[\delta_{n}d,0]^{T}$,
where $\delta_{n}=\frac{2n-N+1}{2}$, $n\in\{0,\dots,N-1\}$. The
steering vector characterizes the spatial phase differences caused
by the varying distances between each antenna element and the receiver
location \cite{xing2023location}. By selecting the $0$-th element
as the reference point, the steering vector of the ULA towards a given
coordinate $\boldsymbol{l}$ is expressed as 
\begin{align}
 & \boldsymbol{v}(\boldsymbol{l})\nonumber \\
= & [1,e^{-j\frac{2\pi}{\lambda}((\|\boldsymbol{l}-\boldsymbol{u}_{1}\|-\|\boldsymbol{l}-\boldsymbol{u}_{0}\|))},\dots,e^{-j\frac{2\pi}{\lambda}((\|\boldsymbol{l}-\boldsymbol{u}_{N-1}\|-\|\boldsymbol{l}-\boldsymbol{u}_{0}\|))}]^{T}.\label{eq:steering}
\end{align}
To facilitate the target sensing, let $r$ represent the distance,
$\theta$ denote the angle of a special point location, and $(r,\theta)$
denote the coordinate in polar coordinates. To this end, we convert
the steering vector from the Cartesian coordinate system in \eqref{eq:steering}
to the polar coordinate system. Define $x=r\cos\theta$ and $y=r\sin\theta$,
the steering vector in the polar coordinate is given as \cite{cui2022channel}
\begin{equation}
\boldsymbol{v}(\theta,r)=[1,\dots,e^{-j\frac{2\pi}{\lambda}(r^{(N-1)}-r^{0})}]^{T},
\end{equation}
where $r^{(n)}$ denotes the distance between the $n$-th element
and the point with
\begin{equation}
r^{(n)}=\sqrt{r^{2}+(\delta_{n}d)^{2}-2\delta_{n}dr\cos\theta}.
\end{equation}
 Subsequently, we define the free-space path loss vector as 
\begin{equation}
\boldsymbol{b}(\boldsymbol{l})=[\frac{\lambda}{4\pi\|\boldsymbol{l}-\boldsymbol{u}_{0}\|},\dots,\frac{\lambda}{4\pi\|\boldsymbol{l}-\boldsymbol{u}_{N-1}\|}]^{T}.
\end{equation}
Let $\boldsymbol{g}_{k}\in\mathbb{C}^{N\times1}$ denote the channel
vector where $k=0$ corresponds to the legitimate CU link and $k\in\mathcal{K}_{\text{EAV}}=\{1,...,K+1\}$
represents potential eavesdroppers' link. We consider that each near-field
eavesdropping channel $\boldsymbol{g}_{k}$ comprises one LoS path
and multi-path NLoS paths. Let $\boldsymbol{g}_{k}^{\mathrm{NLoS}}\in\mathbb{C}^{N\times1}$
denote the NLoS component of the eavesdropping channel $\boldsymbol{g}_{k}$.
We can express the overall channel $\boldsymbol{g}_{k}$ as 
\begin{equation}
\boldsymbol{g}_{k}=\underset{\text{Geometric component}}{\underbrace{\boldsymbol{v}(\boldsymbol{l}_{k})\odot\boldsymbol{b}(\boldsymbol{l}_{k})}}+\boldsymbol{g}_{k}^{\mathrm{NLoS}},\forall k\in\mathcal{K}_{\text{EAV}}.\label{eq:steering-1}
\end{equation}

\subsection{Secure Communication Model}

In this subsection, we consider the secure communication framework
for the near-field secure ISCAP scenario. Recall that $\boldsymbol{g}_{0}\in\mathbb{C}^{N\times1}$
represents the channel vector between the BS and the CU. The received
signal at the CU is expressed as
\begin{equation}
y_{0}(t)=\boldsymbol{g}_{0}^{H}\boldsymbol{w}_{0}s_{0}(t)+\boldsymbol{g}_{0}^{H}\boldsymbol{s}_{1}(t)+z_{0}(t),\label{eq:Received signal at CU}
\end{equation}
where $z_{0}(t)\sim\mathcal{CN}(0,\sigma_{0}^{2})$ denotes the additive
white Gaussian noise (AWGN) at the CU receiver with $\sigma_{0}^{2}$
denoting the noise power. Based on the transmit covariance defined
in \eqref{eq:covariance} and received signal in \eqref{eq:Received signal at CU},
the received \ac{sinr} at the CU is
\begin{equation}
\gamma_{0}(\boldsymbol{R}_{0},\boldsymbol{R}_{1})=\frac{\boldsymbol{g}_{0}^{H}\boldsymbol{R}_{0}\boldsymbol{g}_{0}}{\boldsymbol{g}_{0}^{H}\boldsymbol{R}_{1}\boldsymbol{g}_{0}+\sigma_{0}^{2}}.\label{eq:SINR at CU}
\end{equation}
Furthermore, the received signal at eavesdropper $k\in\mathcal{K}_{\mathrm{EAV}}$
is denoted as
\begin{equation}
y_{k}(t)=\boldsymbol{g}_{k}^{H}\boldsymbol{w}_{0}s_{0}(t)+\boldsymbol{g}_{k}^{H}\boldsymbol{s}_{1}(t)+z_{k}(t),\label{eq:Received signal at target}
\end{equation}
where $z_{k}(t)\sim\mathcal{CN}(0,\sigma_{k}^{2})$ denotes the AWGN
at the receiver of eavesdropper $k\in\mathcal{K}_{\mathrm{EAV}}$
with $\sigma_{k}^{2}$ denoting the noise power. Similarly, the received
SINR at eavesdropper $k\in\mathcal{K}_{\mathrm{EAV}}$ is
\begin{equation}
\gamma_{k}(\boldsymbol{R}_{0},\boldsymbol{R}_{1},\boldsymbol{g}_{k})=\frac{\boldsymbol{g}_{k}^{H}\boldsymbol{R}_{0}\boldsymbol{g}_{k}}{\boldsymbol{g}_{k}^{H}\boldsymbol{R}_{1}\boldsymbol{g}_{k}+\sigma_{k}^{2}}.
\end{equation}

\subsection{Sensing and Powering Models}

Furthermore, we consider energy harvesting at the ERs and target sensing
in the near-field ISCAP scenario. To begin with, the received signal
at ER $k\in\mathcal{K}_{\mathrm{ER}}$ (by omitting the noise) is
given as
\begin{equation}
\boldsymbol{r}_{k}(t)=\boldsymbol{g}_{k}^{H}\boldsymbol{x}(t).\label{eq:Received signal at target-1}
\end{equation}
Since each ER can harvest wireless energy from both information and
dedicated sensing signals, the harvested wireless power (energy-per-unit-time)
at ER $k\in\mathcal{K}_{\mathrm{ER}}$ is given as 
\begin{equation}
E_{k}(\boldsymbol{R}_{0},\boldsymbol{R}_{1},\boldsymbol{g}_{k})=\zeta\boldsymbol{g}_{k}^{H}(\boldsymbol{R}_{0}+\boldsymbol{R}_{1})\boldsymbol{g}_{k},\label{eq:EH}
\end{equation}
where $0\leq\zeta\leq1$ denotes the energy harvesting efficiency.\textcolor{blue}{{}
}\textcolor{black}{Although} we assume a linear energy harvesting
efficiency, our proposed designs can be readily extended to senarios
with non-linear energy harvesting efficiency \cite{clerckx2018fundamentals}.

Next, we discuss the adopted target sensing model. We assume that
the BS aims to perform target localization by analyzing the reflected
echo signals. For near-field target localization, both distance and
angle can be estimated simultaneously exploiting the steering vector,
enabling precise localization based on the near-field echo signals
\cite{qu2023near}. Let $r_{s}$ and $\theta_{s}$ denote the distance
and the angle of the sensing target to the origin, respectively. Let
$\boldsymbol{X}=[\boldsymbol{x}(1),\boldsymbol{x}(2),\dots,\boldsymbol{x}(T)]\in\mathbb{C}^{N\times T}$
and $\boldsymbol{Y}_{s}\in\mathbb{C}^{N\times T}$ denote the accumulated
transmitted signal and received echo signal over the $T$ time slots.
The received echo signal $\boldsymbol{Y}_{s}$ at the BS is denoted
as \cite{lu2023beamforming,qu2023near}\footnote{We consider that self-interference mitigation is performed of the
BS which combines isolation and cancellation to suppress leakage for
achieving full-duplex ISAC. }
\begin{equation}
\boldsymbol{Y}_{s}=\beta_{s}\boldsymbol{v}(\theta_{s},r_{s})\boldsymbol{v}^{T}(\theta_{s},r_{s})\boldsymbol{X}+\boldsymbol{Z}_{s},
\end{equation}
where $\beta_{s}\in\mathbb{C}$ denotes the complex round-trip channel
coefficient associated with the target depending on the path loss
and its radar cross section (RCS), $\boldsymbol{Z}_{s}\in\mathbb{C}^{N\times T}$
denotes the background noise at the BS receiver (including clutter
or interference) with each entry being a zero-mean CSCG random variable
with variance $\sigma_{s}^{2}$. Then, we vectorize matrix $\boldsymbol{Y}_{s}$
as
\begin{equation}
\boldsymbol{y}_{s}=\hat{\boldsymbol{x}}+\hat{\boldsymbol{z}},
\end{equation}
where $\hat{\boldsymbol{x}}=\mathrm{vec}(\beta_{s}\boldsymbol{v}(\theta_{s},r_{s})\boldsymbol{v}^{T}(\theta_{s},r_{s})\boldsymbol{X})$
and $\hat{\boldsymbol{z}}=\mathrm{vec}(\boldsymbol{Z}_{s})$. In this
scenario, we aim to localize the target via estimating $r_{s}$ and
$\theta_{s}$. We denote $\boldsymbol{\xi}=[\theta_{s},r_{s},\mathrm{Re}(\beta_{s}),\mathrm{Im}(\beta_{s})]$
as the unknown parameters to be estimated. The Fisher information
matrix (FIM) $\boldsymbol{J}_{\xi}$ for estimating $\boldsymbol{\xi}$
is given as \cite{hua2023mimo}
\begin{equation}
\boldsymbol{J}_{\xi}[m,n]=\frac{1}{\sigma_{s}^{2}}\mathrm{Re}\Big(\frac{\partial\hat{\boldsymbol{x}}^{H}}{\partial\boldsymbol{\xi}[i]}\frac{\partial\hat{\boldsymbol{x}}}{\partial\boldsymbol{\xi}[j]}\Big),m,n\in\{1,\dots,4\}.
\end{equation}
The CRB matrix is given by the inverse of the FIM and its diagonal
elements correspond to the CRB of parameters to be estimated. For
notational convenience, let $\boldsymbol{A}=\boldsymbol{v}(\theta_{s},r_{s})\boldsymbol{v}^{T}(\theta_{s},r_{s})$,
$\dot{\boldsymbol{A}}_{\theta}=\frac{\partial\boldsymbol{A}}{\partial\theta_{s}}$,
and $\dot{\boldsymbol{A}}_{r}=\frac{\partial\boldsymbol{A}}{\partial r_{s}}$.
According to \cite{qu2023near} the CRB for estimating $\theta_{s}$
is given as
\begin{equation}
\begin{array}{cl}
 & \mathrm{CRB}_{\theta}(\theta_{s},\boldsymbol{R}_{0},\boldsymbol{R}_{1})\\
 & =\frac{\sigma_{s}^{2}}{2|\beta_{s}|^{2}T}\frac{\mathrm{tr}(\boldsymbol{A}\boldsymbol{R}\boldsymbol{A}^{H})}{\mathrm{tr}(\dot{\boldsymbol{A}}_{\theta}\boldsymbol{R}\dot{\boldsymbol{A}}_{\theta}^{H})\mathrm{tr}(\boldsymbol{A}\boldsymbol{R}\boldsymbol{A}^{H})-|\mathrm{tr}(\boldsymbol{A}\boldsymbol{R}\dot{\boldsymbol{A}}_{\theta}^{H})|^{2}},
\end{array}\label{eq:CRB1_theta}
\end{equation}
in which $\boldsymbol{R}=\boldsymbol{R}_{0}+\boldsymbol{R}_{1}$.
Similarly, the CRB for estimating $r_{s}$ is given as
\begin{equation}
\begin{array}{cl}
 & \mathrm{CRB}_{r}(r_{s},\boldsymbol{R}_{0},\boldsymbol{R}_{1})\\
 & =\frac{\sigma_{s}^{2}}{2|\beta_{s}|^{2}T}\frac{\mathrm{tr}(\boldsymbol{A}\boldsymbol{R}\boldsymbol{A}^{H})}{\mathrm{tr}(\dot{\boldsymbol{A}}_{r}\boldsymbol{R}\dot{\boldsymbol{A}}_{r}^{H})\mathrm{tr}(\boldsymbol{A}\boldsymbol{R}\boldsymbol{A}^{H})-|\mathrm{tr}(\boldsymbol{A}\boldsymbol{R}\dot{\boldsymbol{A}}_{r}^{H})|^{2}}.
\end{array}\label{eq:CRB_r}
\end{equation}

\subsection{Near-field CSI Uncertainty Models }

In this subsection, we introduce the near-field channel uncertainty
model. Regarding the CU channel $\boldsymbol{g}_{0}\in\mathbb{C}^{N\times1}$,
we assume that the BS can acquire accurate CSI through standard channel
estimation techniques \cite{cui2022channel}. In contrast, we adopt
a location-based uncertain CSI model for the ERs and the sensing target,
which act as potential eavesdroppers. The technical foundation lies
in the inherent relationship between geometric location and electromagnetic
wave propagation characteristics in modern networks \cite{su2023sensing}.
In general, the BS can obtain approximate locations of targets and
ERs through its sensing capabilities. 

Therefore, in our paper, we assume that the BS has access only to
coarse location information for all potential eavesdroppers, which
is denoted as $\{\hat{\boldsymbol{l}_{k}}|\hat{\boldsymbol{l}_{k}}=[\hat{x_{k}},\hat{y_{k}}]^{T},k\in\mathcal{K}_{\mathrm{EAV}}\}$.
The actual location (ground truth) of eavesdroppers' is denoted as
$\{\boldsymbol{l}_{k}|\boldsymbol{l}_{k}=[x_{k},y_{k}]^{T},k\in\mathcal{K}_{\mathrm{EAV}}\}$.
Due to practical limitations in measurement accuracy, the estimated
coarse location of the eavesdropper inherently contains uncertainty.
Therefore, the actual location $\boldsymbol{l}_{k}$ can be modeled
as the estimated location $\hat{\boldsymbol{l}_{k}}$ plus a bounded
error term $\Delta\boldsymbol{l}_{k}$, which has a known bound $\varepsilon_{k}$
and is determined by the positioning accuracy \cite{xu2022robust,xing2023location},
i.e., 
\begin{equation}
\Delta\boldsymbol{l}_{k}=\boldsymbol{l}_{k}-\hat{\boldsymbol{l}_{k}},\|\Delta\boldsymbol{l}_{k}\|\leq\varepsilon_{k}.\label{eq:location bound}
\end{equation}

Next, we introduce the near-field eavesdropping channel uncertainty
model. Recent channel measurement studies have revealed critical insights
into near-field propagation scenarios \cite{cui2022channel}. These
studies indicate that while a limited number of NLoS signal paths
may reach the receiver, the LoS path dominates the received power.
Specifically, the LoS path has been observed to exhibit power levels
approximately $13$ dB higher than the combined power of all NLoS
components, contributing to majority of the total channel power \cite{cui2022channel,Wang2020IRS,cui2022near}.
As a result, we characterize the NLoS component by a bounded norm
$\delta_{k}$ \cite{xu2022robust,xing2023location}, i.e., 
\begin{equation}
\|\boldsymbol{g}_{k}^{\mathrm{NLoS}}\|\leq\delta_{k}.
\end{equation}
In this case, based on the location error and NLoS channel component,
the eavesdropping channel uncertainty for potential eavesdropper $k\in\mathcal{K}_{\mathrm{EAV}}$
is characterized by a CSI error $\Delta\boldsymbol{g}_{k}$ as
\begin{eqnarray}
\Delta\boldsymbol{g}_{k} & = & \boldsymbol{g}_{k}-\hat{\boldsymbol{g}_{k}}\nonumber \\
 & = & \underset{\text{Geometric uncertainty}}{\underbrace{\boldsymbol{v}(\boldsymbol{l}_{k})\odot\boldsymbol{b}(\boldsymbol{l}_{k})-\boldsymbol{v}(\hat{\boldsymbol{l}_{k}})\odot\boldsymbol{b}(\hat{\boldsymbol{l}_{k}})}}+\boldsymbol{g}_{k}^{\mathrm{NLoS}}.\label{eq:uncertain region}
\end{eqnarray}
As a result, we denote the eavesdropping channel uncertainty region
of potential eavesdropper $k\in\mathcal{K}_{\mathrm{EAV}}$ as $\Psi_{k}$,
where
\begin{equation}
\Psi_{k}=\{\boldsymbol{g}_{k}|\boldsymbol{g}_{k}=\hat{\boldsymbol{g}_{k}}+\Delta\boldsymbol{g}_{k},\|\Delta\boldsymbol{l}_{k}\|\leq\varepsilon_{k},\|\boldsymbol{g}_{k}^{\mathrm{NLoS}}\|\leq\delta_{k}\}.\label{eq:uncertain region-1}
\end{equation}

\subsection{Problem Formulation}

In this subsection, we define the worst-case performance metrics for
secure communication, sensing, and powering, and subsequently formulate
a robust joint beamforming problem. 

To begin with, we consider the worst-case achievable secrecy rate
at the CU based on the secure communication model, which is given
by \cite{HuangLeeJ12}
\begin{equation}
\begin{aligned}C(\boldsymbol{R}_{0},\boldsymbol{R}_{1}) & =\underset{\boldsymbol{g}_{k}\in\Psi_{k},k\in\mathcal{K}_{\mathrm{EAV}}}{\min}\Big(\log_{2}\big(1+\gamma_{0}(\boldsymbol{R}_{0},\boldsymbol{R}_{1})\big)\\
 & -\log_{2}\big(1+\gamma_{k}(\boldsymbol{R}_{0},\boldsymbol{R}_{1},\boldsymbol{g}_{k})\big)\Big)^{+}.
\end{aligned}
\label{eq:rate}
\end{equation}
Similarly, based on \eqref{eq:EH}, we consider the worst-case energy
harvesting performance at each ER $k\in\mathcal{K}_{\mathrm{ER}}$
as 
\begin{equation}
\hat{E_{k}}(\boldsymbol{R}_{0},\boldsymbol{R}_{1},\boldsymbol{g}_{k})=\underset{\boldsymbol{g}_{k}\in\Psi_{k}}{\min}\zeta\boldsymbol{g}_{k}^{H}(\boldsymbol{R}_{0}+\boldsymbol{R}_{1})\boldsymbol{g}_{k}.
\end{equation}

For target sensing, since the exact location of the target is unknown,
it is necessary for the system design to ensure worst-case performance
under a predefined uncertainty region. Let the uncertainty region
of the target be denoted as 

\begin{equation}
\Theta=\{(\theta_{s},r_{s})|\theta_{s}\in[\theta_{s}^{\textrm{L}},\theta_{s}^{\textrm{U}}],\theta_{s}\in[r_{s}^{\textrm{L}},r_{s}^{\textrm{U}}]\}.
\end{equation}
where $\theta_{s}^{\textrm{L}}$ and $\theta_{s}^{\textrm{U}}$ denote
the lower and upper bounds for the angle uncertainty, respectively,
and $r_{s}^{\textrm{L}}$ and $r_{s}^{\textrm{U}}$ denote the lower
and upper bounds for the distance uncertainty, respectively. However,
ensuring robust sensing performance across the entire uncertainty
region is often challenging. To address this challenge, we employ
a discrete sampling approach within the target's uncertainty region
as in \cite{su2020secure,cheng2024optimal} to approximate and guarantee
the worst-case performance. Let $\bar{\Theta}$ denote the set for
$M$ mesh grid sampling in the uncertain region $\Theta$, i.e., 
\begin{equation}
\begin{array}{r}
\bar{\Theta}=\big\{(\theta_{m},r_{m})|\theta_{m}\in[\theta_{s}^{\textrm{L}},\theta_{s}^{\textrm{U}}],r_{m}\in[r_{s}^{\textrm{L}},r_{s}^{\textrm{U}}],m=1,\dots,M\}.\end{array}
\end{equation}
Then, the worst-case CRB guarantee constraints in the uncertainty
region is given as 
\begin{equation}
\begin{array}{r}
\mathrm{CRB}_{\theta}(\theta_{m},\boldsymbol{R}_{0},\boldsymbol{R}_{1})\leq\Gamma_{\theta},\mathrm{CRB}_{r}(r_{m},\boldsymbol{R}_{0},\boldsymbol{R}_{1})\leq\Gamma_{r},\\
\forall(\theta_{m},r_{m})\in\bar{\Theta}.
\end{array}
\end{equation}

Our objective is to maximize the worst-case secrecy rate in \eqref{eq:rate},
by jointly optimizing the transmit information covariance matrix $\boldsymbol{R}_{0}$
and the sensing/energy/AN covariance matrix $\boldsymbol{R}_{1}$,
subject to constraints on target sensing and wireless energy harvesting.
The secrecy rate maximization problem is formulated as 
\begin{subequations}
\hspace{-0.6cm}
\begin{eqnarray}
\hspace{-0.6cm}\textrm{(P1):} & \hspace{-0.3cm}\underset{\boldsymbol{R}_{0},\boldsymbol{R}_{1}}{\max} & C(\boldsymbol{R}_{0},\boldsymbol{R}_{1})\nonumber \\
 & \hspace{-0.3cm}\textrm{s.t.} & \mathrm{CRB}_{\theta}(\theta_{s},\boldsymbol{R}_{0},\boldsymbol{R}_{1})\leq\Gamma_{\theta},\forall(\theta_{s},r_{s})\in\Theta,\label{eq:C0}\\
 &  & \mathrm{CRB}_{r}(r_{s},\boldsymbol{R}_{0},\boldsymbol{R}_{1})\leq\Gamma_{r},\forall(\theta_{s},r_{s})\in\Theta,\label{eq:C1}\\
 &  & \underset{\boldsymbol{g}_{k}\in\Psi_{k}}{\min}\zeta\boldsymbol{g}_{k}^{H}(\boldsymbol{R}_{0}+\boldsymbol{R}_{1})\boldsymbol{g}_{k}\geq Q,\forall k\in\mathcal{K}_{\mathrm{ER}},\label{eq:C2}\\
 &  & \mathrm{Tr}(\boldsymbol{R}_{0}+\boldsymbol{R}_{1})\leq P,\label{eq:C3}\\
 &  & \boldsymbol{R}_{0}\succeq\boldsymbol{0},\boldsymbol{R}_{1}\succeq\boldsymbol{0},\label{eq:C4}\\
 &  & \textrm{rank}(\boldsymbol{R}_{0})\le1,\label{eq:C5}
\end{eqnarray}
\end{subequations}
where $\Gamma_{\theta}$, $\Gamma_{r}$, and $Q$ denote the given
thresholds for angle estimation, range estimation, and energy harvesting,
respectively. Solving problem (P1) is generally challenging as the
objective function of secrecy rate is non-concave and the rank constraint
is non-convex. Also, characterizing the uncertain region of eavesdropping
channels $\boldsymbol{g}_{k}$ poses significant challenges. In Section
III, we employ relevant optimization techniques such as SDR, fractional
optimization, and Schur complement to reformulate the problem into
a more tractable form for efficient solution development. 

\section{Secure Joint Beamforming for Solving Problem (P1)}

In this section, we develop a high-quality solution to the formulated
problem (P1). First, we employ advanced optimization techniques to
reformulate (P1) into a more tractable. Then, we derive a rigious
error bound to handle the constraints associated with CSI uncertainty. 

\subsection{Problem Reformulation}

In this subsection, we first apply SDR to tackle the rank constraint
in \eqref{eq:C5}. Note that, the rank constraint in \eqref{eq:C5}
is inherently non-convex. To address this issue, we relax the rank
constraint in \eqref{eq:C5}, yielding the following SDR version of
problem (P1). 
\begin{subequations}
\begin{eqnarray*}
\textrm{(SDR1):} & \underset{\boldsymbol{R}_{0},\boldsymbol{R}_{1}}{\max} & C(\boldsymbol{R}_{0},\boldsymbol{R}_{1})\\
 & \textrm{s.t.} & \mathrm{CRB}_{\theta}(\theta_{m},\boldsymbol{R}_{0},\boldsymbol{R}_{1})\leq\Gamma_{\theta},\forall(\theta_{m},r_{m})\in\Theta,\\
 &  & \mathrm{CRB}_{r}(r_{m},\boldsymbol{R}_{0},\boldsymbol{R}_{1})\leq\Gamma_{r},\forall(\theta_{m},r_{m})\in\Theta,\\
 &  & \underset{\boldsymbol{g}_{k}\in\Psi_{k}}{\min}\zeta\boldsymbol{g}_{k}^{H}(\boldsymbol{R}_{0}+\boldsymbol{R}_{1})\boldsymbol{g}_{k}\geq Q,\forall k\in\mathcal{K}_{\mathrm{ER}},\\
 &  & \mathrm{Tr}(\boldsymbol{R}_{0}+\boldsymbol{R}_{1})\leq P,\\
 &  & \boldsymbol{R}_{0}\succeq\boldsymbol{0},\boldsymbol{R}_{1}\succeq\boldsymbol{0}.
\end{eqnarray*}
\end{subequations}
It is important to note that the SDR approach often yields high-rank
solutions that fail to satisfy the original rank constraint specified
in \eqref{eq:C5}. Consequently, additional processing steps are necessary
to obtain a rank-one solution that either exactly matches or approximates
the original problem's requirements. Specifically, we can construct
an equivalent rank-one solution in our paper, as will be shown in
Proposition 3 later. 

Then, we leverage the Schur complement theorem to transform the worst-case
CRB constraints in \eqref{eq:C0} into linear matrix inequalities
(LMIs), significantly improving computational efficiency. The CRB
constraint $\mathrm{CRB}_{\theta}(\theta_{m},\boldsymbol{R}_{0},\boldsymbol{R}_{1})\leq\Gamma_{\theta}$
in \eqref{eq:C0} is equivalently reformulated as \cite{liu2021cramer,qu2023near}

\begin{equation}
\begin{array}{r}
\left[\begin{array}{cc}
\big(\mathrm{tr}(\dot{\boldsymbol{A}}_{\theta}\boldsymbol{R}\dot{\boldsymbol{A}}_{\theta}^{H})-\frac{\sigma_{s}^{2}}{2|\beta_{s}|^{2}T\Gamma_{\theta}}\big) & \mathrm{tr}(\dot{\boldsymbol{A}_{\theta}}\boldsymbol{R}\boldsymbol{A}^{H})\\
\mathrm{tr}(\boldsymbol{A}\boldsymbol{R}_{}\dot{\boldsymbol{A}}_{\theta}^{H}) & \mathrm{tr}(\boldsymbol{A}\boldsymbol{R}_{}\boldsymbol{A}^{H})
\end{array}\right]\succeq\boldsymbol{0},\\
\forall(\theta_{m},r_{m})\in\Theta.
\end{array}\label{eq:C6}
\end{equation}
 Similarly, the CRB constraint $\mathrm{CRB}_{r}(r_{m},\boldsymbol{R}_{0},\boldsymbol{R}_{1})\leq\Gamma_{r}$
in \eqref{eq:C1} can be reformulated as 
\begin{equation}
\begin{array}{r}
\left[\begin{array}{cc}
\big(\mathrm{tr}(\dot{\boldsymbol{A}}_{r}\boldsymbol{R}\dot{\boldsymbol{A}}_{r}^{H})-\frac{\sigma_{s}^{2}}{2|\beta_{s}|^{2}T\Gamma_{r}}\big) & \mathrm{tr}(\dot{\boldsymbol{A}_{r}}\boldsymbol{R}_{}\boldsymbol{A}^{H})\\
\mathrm{tr}(\boldsymbol{A}\boldsymbol{R}_{}\dot{\boldsymbol{A}}_{r}^{H}) & \mathrm{tr}(\boldsymbol{A}\boldsymbol{R}_{}\boldsymbol{A}^{H})
\end{array}\right]\succeq\boldsymbol{0},\\
\forall(\theta_{m},r_{m})\in\Theta.
\end{array}\label{eq:C7-1}
\end{equation}

Next, we tackle the non-concave worst-case secrecy rate objective
function. First, we introduce an auxiliary optimization variable $\gamma_{\mathrm{R}}$
for eavesdropping SINR threshold and we can equivalently reformulate
problem (SDR1) as
\begin{alignat*}{1}
\textrm{(P1.1):}\underset{\boldsymbol{R}_{0},\boldsymbol{R}_{1},\gamma_{\mathrm{R}}}{\max} & \frac{\boldsymbol{g}_{0}^{H}\boldsymbol{R}_{0}\boldsymbol{g}_{0}}{\boldsymbol{g}_{0}^{H}\boldsymbol{R}_{1}\boldsymbol{g}_{0}+\sigma_{0}^{2}}\\
\textrm{s.t.} & \underset{\boldsymbol{g}_{k}\in\Psi_{k}}{\max}\frac{\boldsymbol{g}_{k}^{H}\boldsymbol{R}_{0}\boldsymbol{g}_{k}}{\boldsymbol{g}_{k}^{H}\boldsymbol{R}_{1}\boldsymbol{g}_{k}+\sigma_{k}^{2}}\leq\gamma_{\mathrm{R}},\forall k\in\mathcal{K}_{\mathrm{EAV}}\\
 & \underset{\boldsymbol{g}_{k}\in\Psi_{k}}{\min}\zeta\boldsymbol{g}_{k}^{H}(\boldsymbol{R}_{0}+\boldsymbol{R}_{1})\boldsymbol{g}_{k}\geq Q,\forall k\in\mathcal{K}_{\mathrm{ER}},\\
 & \mathrm{Tr}(\boldsymbol{R}_{0}+\boldsymbol{R}_{1})\leq P,\\
 & \boldsymbol{R}_{0}\succeq\boldsymbol{0},\boldsymbol{R}_{1}\succeq\boldsymbol{0},\\
 & \textrm{\eqref{eq:C6} and \eqref{eq:C7-1}}.
\end{alignat*}

Subsequently, we utilize fractional optimization to reformulate the
objective function into a simplified form. Specifically, we introduce
a slack optimization variable $\xi>0$ to apply the Charnes-Cooper
transformation \cite{shen2018fractional}. To begin with, let $\hat{\boldsymbol{R}_{0}}=\xi\boldsymbol{R}_{0}$
and $\hat{\boldsymbol{R}_{1}}=\xi\boldsymbol{R}_{1}$. Based on this
transformation, we further update the CRB constraints in \eqref{eq:C6}
and \eqref{eq:C7-1} as 
\begin{equation}
\begin{array}{r}
\left[\begin{array}{cc}
\big(\mathrm{tr}(\dot{\boldsymbol{A}}_{\theta}\hat{\boldsymbol{R}}\dot{\boldsymbol{A}}_{\theta}^{H})-\frac{\xi\sigma_{s}^{2}}{2|\beta_{s}|^{2}T\Gamma_{\theta}}\big) & \mathrm{tr}(\dot{\boldsymbol{A}_{\theta}}\hat{\boldsymbol{R}}\boldsymbol{A}^{H})\\
\mathrm{tr}(\boldsymbol{A}\hat{\boldsymbol{R}}\dot{\boldsymbol{A}}_{\theta}^{H}) & \mathrm{tr}(\boldsymbol{A}\hat{\boldsymbol{R}}\boldsymbol{A}^{H})
\end{array}\right]\succeq\boldsymbol{0},\\
\forall(\theta_{m},r_{m})\in\Theta,
\end{array}\label{eq:CRB1-1}
\end{equation}
\begin{equation}
\begin{array}{r}
\left[\begin{array}{cc}
\big(\mathrm{tr}(\dot{\boldsymbol{A}}_{r}\hat{\boldsymbol{R}}\dot{\boldsymbol{A}}_{r}^{H})-\frac{\xi\sigma_{s}^{2}}{2|\beta_{s}|^{2}T\Gamma_{r}}\big) & \mathrm{tr}(\dot{\boldsymbol{A}_{r}}\hat{\boldsymbol{R}}\boldsymbol{A}^{H})\\
\mathrm{tr}(\boldsymbol{A}\hat{\boldsymbol{R}}\dot{\boldsymbol{A}}_{r}^{H}) & \mathrm{tr}(\boldsymbol{A}\hat{\boldsymbol{R}}\boldsymbol{A}^{H})
\end{array}\right]\succeq\boldsymbol{0},\\
\forall(\theta_{m},r_{m})\in\Theta,
\end{array}\label{eq:CRB2-1}
\end{equation}
respectively, where $\hat{\boldsymbol{R}}=\hat{\boldsymbol{R}_{0}}+\hat{\boldsymbol{R}_{1}}$.
Similarly, the worst-case eavesdropping SINR constraints in problem
(P1.1), the worst-case harvested power constraints in \eqref{eq:C2},
and the transmit power constraint in \eqref{eq:C3} are respectively
modified as 
\begin{eqnarray}
 &  & \underset{\boldsymbol{g}_{k}\in\Psi_{k}}{\max}\frac{\boldsymbol{g}_{k}^{H}\hat{\boldsymbol{R}_{0}}\boldsymbol{g}_{k}}{\boldsymbol{g}_{k}^{H}\hat{\boldsymbol{R}_{1}}\boldsymbol{g}_{k}+\xi\sigma_{k}^{2}}\leq\gamma_{\mathrm{R}},\forall k\in\mathcal{K}_{\mathrm{EAV}},\label{eq:C7}\\
 &  & \underset{\boldsymbol{g}_{k}\in\Psi_{k}}{\min}\zeta\boldsymbol{g}_{k}^{H}(\hat{\boldsymbol{R}_{0}}+\hat{\boldsymbol{R}_{1}})\boldsymbol{g}_{k}\geq\xi Q,\forall k\in\mathcal{K}_{\mathrm{ER}},\label{eq:C8}\\
 &  & \mathrm{Tr}(\hat{\boldsymbol{R}_{0}}+\hat{\boldsymbol{R}_{1}})\leq\xi P.\label{eq:C9}
\end{eqnarray}
As a result, we can effectively recast problem (P1.1) into the following
equivalent form that is more tractable for analysis and optimization:
\begin{eqnarray*}
\textrm{(P2):} & \underset{\hat{\boldsymbol{R}_{0}},\hat{\boldsymbol{R}_{1}},\gamma_{\mathrm{R}},\xi>0}{\max} & \boldsymbol{g}_{0}^{H}\hat{\boldsymbol{R}_{0}}\boldsymbol{g}_{0}\\
 & \textrm{s.t.} & \hspace{-0.8cm}\underset{\boldsymbol{g}_{k}\in\Psi_{k}}{\max}\frac{\boldsymbol{g}_{k}^{H}\hat{\boldsymbol{R}_{0}}\boldsymbol{g}_{k}}{\boldsymbol{g}_{k}^{H}\hat{\boldsymbol{R}_{1}}\boldsymbol{g}_{k}+\xi\sigma_{k}^{2}}\leq\gamma_{\mathrm{R}},\forall k\in\mathcal{K}_{\mathrm{EAV}},\\
 &  & \hspace{-0.8cm}\underset{\boldsymbol{g}_{k}\in\Psi_{k}}{\min}\zeta\boldsymbol{g}_{k}^{H}(\hat{\boldsymbol{R}_{0}}+\hat{\boldsymbol{R}_{1}})\boldsymbol{g}_{k}\geq\xi Q,\forall k\in\mathcal{K}_{\mathrm{ER}},\\
 &  & \hspace{-0.8cm}\boldsymbol{g}_{0}^{H}\hat{\boldsymbol{R}_{1}}\boldsymbol{g}_{0}+\xi\sigma_{0}^{2}=1,\\
 &  & \hspace{-0.8cm}\mathrm{Tr}(\hat{\boldsymbol{R}_{0}}+\hat{\boldsymbol{R}_{1}})\leq\xi P,\\
 &  & \hspace{-0.8cm}\hat{\boldsymbol{R}_{0}}\succeq\boldsymbol{0},\hat{\boldsymbol{R}_{1}}\succeq\boldsymbol{0},\\
 &  & \hspace{-0.8cm}\textrm{\eqref{eq:CRB1-1} and \eqref{eq:CRB2-1}}.
\end{eqnarray*}
Note that the worst-case eavesdropping SINR threshold $\gamma_{\mathrm{R}}$
appears exclusively in \eqref{eq:C7} and is intricately coupled with
other variables, rendering it mathematically challenging to analyze
and handle effectively. To solve problem (P2), we adopt a two-step
optimization approach. First, we address problem (P2) for a fixed
value $\gamma_{\mathrm{R}}$ by leveraging its specific structure
to obtain an efficient solution. The reformulated problem is given
as
\begin{eqnarray*}
\textrm{(P3):} & \underset{\hat{\boldsymbol{R}_{0}},\hat{\boldsymbol{R}_{1}},\xi>0}{\max} & \boldsymbol{g}_{0}^{H}\hat{\boldsymbol{R}_{0}}\boldsymbol{g}_{0}\\
 & \textrm{s.t.} & \boldsymbol{g}_{0}^{H}\hat{\boldsymbol{R}_{1}}\boldsymbol{g}_{0}+\xi\sigma_{0}^{2}=1,\\
 &  & \hat{\boldsymbol{R}_{0}}\succeq\boldsymbol{0},\hat{\boldsymbol{R}_{1}}\succeq\boldsymbol{0},\\
 &  & \textrm{\eqref{eq:CRB1-1}, \eqref{eq:CRB2-1}, \eqref{eq:C7}, \eqref{eq:C8}, and \eqref{eq:C9}. }
\end{eqnarray*}
 Subsequently, we perform a 1D search over $\gamma_{\mathrm{R}}$
to identify the optimal $\gamma_{\mathrm{R}}$ that minimizes the
objective of (P2). 

It should be noted that problem (P3) remains highly challenging to
solve due to the worst-case eavesdropping SINR constraints in \eqref{eq:C7}
and the worst-case energy harvesting constraints in \eqref{eq:C8}.
In fact, these constraints are difficult to express explicitly in
closed-form representations. In the following subsections, we propose
to solve problem (P3) based on CSI error determination for the eavesdropping
CSI uncertainty region.

\subsection{Solution to Problem (P3) based on CSI Error Bound Determination}

In this subsection, we first derive the CSI error bounds for a given
eavesdropping CSI uncertainty region characterized in \eqref{eq:uncertain region},
which depends on the eavesdropper's location. Subsequently, we employ
robust optimization techniques based on the S-Procedure to propose
a solution to problem (P3) under the estiblished CSI error bounds. 

To begin with, we consider the CSI error bound for potential eavesdropper
$k\in\mathcal{K}_{\mathrm{EAV}}$ as
\begin{eqnarray}
\|\Delta\boldsymbol{g}_{k}\| & = & \|\boldsymbol{g}_{k}-\hat{\boldsymbol{g}_{k}}\|\nonumber \\
 & = & \|\boldsymbol{v}(\boldsymbol{l}_{k})\odot\boldsymbol{b}(\boldsymbol{l}_{k})-\boldsymbol{v}(\hat{\boldsymbol{l}_{k}})\odot\boldsymbol{b}(\hat{\boldsymbol{l}_{k}})+\boldsymbol{g}_{k}^{\mathrm{NLoS}}\|\nonumber \\
 & \overset{(a)}{\leq} & \|\boldsymbol{v}(\boldsymbol{l}_{k})\odot\boldsymbol{b}(\boldsymbol{l}_{k})-\boldsymbol{v}(\hat{\boldsymbol{l}_{k}})\odot\boldsymbol{b}(\hat{\boldsymbol{l}_{k}})\|+\delta_{k},
\end{eqnarray}
where inequality (a) follows from the triangle inequality and the
term $\|\boldsymbol{v}(\boldsymbol{l}_{k})\odot\boldsymbol{b}(\boldsymbol{l}_{k})-\boldsymbol{v}(\hat{\boldsymbol{l}_{k}})\odot\boldsymbol{b}(\hat{\boldsymbol{l}_{k}})\|$
represents the CSI LoS component error caused by the location uncertainty
of the potential eavesdropper $k\in\mathcal{K}_{\mathrm{EAV}}$. 

Next, we proceed to derive an explicit upper bound for the LoS-component
CSI error. Recall that the Cartesian coordinate vector of its $n$-th
antenna element is denoted as $\boldsymbol{u}_{n}=[\delta_{n}d,0]^{T}$,
where $\delta_{n}=\frac{2n-N+1}{2}$, $n\in\{0,\dots,N-1\}$. The
expression of the LoS-component CSI error is provided in \eqref{eq:oribound}
at the top of next page. For clarity, we introduce a refined definition
of $\Upsilon_{k}(\Delta\boldsymbol{l}_{k})$ within the same equation.
The primary objective is to determine the CSI error bound concerning
$\Delta\boldsymbol{l}_{k}$ under the constraint $\|\Delta\boldsymbol{l}_{k}\|\leq\varepsilon_{k}$
in \eqref{eq:location bound}. This task is inherently challenging
due to the non-linear appearance of $\Delta\boldsymbol{l}_{k}$ in
both the denominator and the cosine function, which introduces significant
mathematical complexity and hinders the straightforward derivation
of an explicit upper bound. To address this and motivated by \cite{xing2023location},
we leverage Taylor series expansion techniques to approximate the
upper bound while minimizing computational overhead. It is evident
that approximating the upper bound for $\Upsilon_{k}(\Delta\boldsymbol{l}_{k})$
is up to determining the two distinct components: $\varPi_{k}(\Delta\boldsymbol{l}_{k})$
and $\varOmega_{k}(\Delta\boldsymbol{l}_{k})$ defined in \eqref{eq:bound1}
and \eqref{eq:bound2-1}, respectively. The first component involves
a non-linear term in the denominator, while the second combines a
linear denominator with a cosine function, further complicating the
analytical process. In the subsequent analysis, we derive the upper
bounds for these two components individually. 
\begin{figure*}
\centering 
\begin{eqnarray}
 &  & \|\boldsymbol{v}(\boldsymbol{l}_{k})\odot\boldsymbol{b}(\boldsymbol{l}_{k})-\boldsymbol{v}(\hat{\boldsymbol{l}_{k}})\odot\boldsymbol{b}(\hat{\boldsymbol{l}_{k}})\|\nonumber \\
 & = & \frac{\lambda}{4\pi}\sqrt{\sum_{n=0}^{N-1}\Bigg(\frac{e^{-j\frac{2\pi}{\lambda}((\|\boldsymbol{l}_{k}-\boldsymbol{u}_{n}\|-\|\boldsymbol{l}_{k}-\boldsymbol{u}_{0}\|))}}{\|\boldsymbol{l}_{k}-\boldsymbol{u}_{n}\|}-\frac{e^{-j\frac{2\pi}{\lambda}((\|\hat{\boldsymbol{l}_{k}}-\boldsymbol{u}_{n}\|-\|\hat{\boldsymbol{l}_{k}}-\boldsymbol{u}_{0}\|))}}{\|\hat{\boldsymbol{l}_{k}}-\boldsymbol{u}_{n}\|}\Bigg)^{2}}=\frac{\lambda}{4\pi}\sqrt{\Upsilon_{k}(\Delta\boldsymbol{l}_{k})},\label{eq:oribound}\\
\Upsilon_{k}(\Delta\boldsymbol{l}_{k}) & = & \varPi_{k}(\Delta\boldsymbol{l}_{k})+\varOmega_{k}(\Delta\boldsymbol{l}_{k})\\
 & = & \sum_{n=0}^{N-1}\Bigg(\frac{1}{\|\hat{\boldsymbol{l}_{k}}+\Delta\boldsymbol{l}_{k}-\boldsymbol{u}_{n}\|^{2}}+\frac{1}{\|\hat{\boldsymbol{l}_{k}}-\boldsymbol{u}_{n}\|^{2}}-\nonumber \\
 &  & \frac{2}{\|\hat{\boldsymbol{l}_{k}}+\Delta\boldsymbol{l}_{k}-\boldsymbol{u}_{n}\|\|\hat{\boldsymbol{l}_{k}}-\boldsymbol{u}_{n}\|}\cos\Big(\frac{2\pi\big(\|\hat{\boldsymbol{l}_{k}}+\Delta\boldsymbol{l}_{k}-\boldsymbol{u}_{n}\|-\|\hat{\boldsymbol{l}_{k}}+\Delta\boldsymbol{l}_{k}-\boldsymbol{u}_{0}\|-\|\hat{\boldsymbol{l}_{k}}-\boldsymbol{u}_{n}\|+\|\hat{\boldsymbol{l}_{k}}-\boldsymbol{u}_{0}\|\big)}{\lambda}\Big)\Bigg).\nonumber \\
\varPi_{k}(\Delta\boldsymbol{l}_{k}) & = & \sum_{n=0}^{N-1}\frac{1}{\|\hat{\boldsymbol{l}_{k}}+\Delta\boldsymbol{l}_{k}-\boldsymbol{u}_{n}\|^{2}}\label{eq:bound1}\\
\varOmega_{k}(\Delta\boldsymbol{l}_{k}) & = & -2\sum_{n=0}^{N-1}\frac{\cos\Big(\frac{2\pi\big(\|\hat{\boldsymbol{l}_{k}}+\Delta\boldsymbol{l}_{k}-\boldsymbol{u}_{n}\|-\|\hat{\boldsymbol{l}_{k}}+\Delta\boldsymbol{l}_{k}-\boldsymbol{u}_{0}\|-\|\hat{\boldsymbol{l}_{k}}-\boldsymbol{u}_{n}\|+\|\hat{\boldsymbol{l}_{k}}-\boldsymbol{u}_{0}\|\big)}{\lambda}\Big)}{\|\hat{\boldsymbol{l}_{k}}-\boldsymbol{u}_{n}\|\|\hat{\boldsymbol{l}_{k}}+\Delta\boldsymbol{l}_{k}-\boldsymbol{u}_{n}\|}.\label{eq:bound2-1}
\end{eqnarray}

\hrule
\end{figure*}

To begin with, we define $\varPi_{k}^{(n)}(\Delta\boldsymbol{l}_{k})$
as 
\begin{equation}
\varPi_{k}^{(n)}(\Delta\boldsymbol{l}_{k})=\frac{1}{\|\hat{\boldsymbol{l}_{k}}+\Delta\boldsymbol{l}_{k}-\boldsymbol{u}_{n}\|^{2}}.
\end{equation}
Applying the triangle inequality, we have $\|\hat{\boldsymbol{l}_{k}}+\Delta\boldsymbol{l}_{k}-\boldsymbol{u}_{n}\|\geq\|\hat{\boldsymbol{l}_{k}}-\boldsymbol{u}_{n}\|-\|\Delta\boldsymbol{l}_{k}\|$.
Then, we have
\begin{eqnarray}
\varPi_{k}^{(n)}(\Delta\boldsymbol{l}_{k}) & = & \frac{1}{\|\hat{\boldsymbol{l}_{k}}+\Delta\boldsymbol{l}_{k}-\boldsymbol{u}_{n}\|^{2}},\nonumber \\
 & \leq & \frac{1}{(\|\hat{\boldsymbol{l}_{k}}-\boldsymbol{u}_{n}\|-\|\Delta\boldsymbol{l}_{k}\|)^{2}}\nonumber \\
 & \overset{(b)}{\leq} & \frac{1}{(\|\hat{\boldsymbol{l}_{k}}-\boldsymbol{u}_{n}\|-\varepsilon_{k})^{2}},\label{eq:bound 1}
\end{eqnarray}
where inequality (b) holds due to $\|\Delta\boldsymbol{l}_{k}\|\leq\varepsilon_{k}$.
As a result, we establish an upper bound for $\varPi_{k}^{(n)}(\Delta\boldsymbol{l}_{k})$
as
\begin{equation}
\varPi_{k}(\Delta\boldsymbol{l}_{k})\leq\sum_{n=0}^{N-1}\frac{1}{(\|\hat{\boldsymbol{l}_{k}}-\boldsymbol{u}_{n}\|-\varepsilon_{k})^{2}}.
\end{equation}

For $\varOmega_{k}(\Delta\boldsymbol{l}_{k})$, we define $\varOmega_{k}^{(n)}(\Delta\boldsymbol{l}_{k})$
as 
\begin{equation}
\begin{array}{l}
\varOmega_{k}^{(n)}(\Delta\boldsymbol{l}_{k})\\
=-\frac{\cos\Big(\frac{2\pi\big(\|\hat{\boldsymbol{l}_{k}}+\Delta\boldsymbol{l}_{k}-\boldsymbol{u}_{n}\|-\|\hat{\boldsymbol{l}_{k}}+\Delta\boldsymbol{l}_{k}-\boldsymbol{u}_{0}\|-\|\hat{\boldsymbol{l}_{k}}-\boldsymbol{u}_{n}\|+\|\hat{\boldsymbol{l}_{k}}-\boldsymbol{u}_{0}\|\big)}{\lambda}\Big)}{\|\hat{\boldsymbol{l}_{k}}-\boldsymbol{u}_{n}\|\|\hat{\boldsymbol{l}_{k}}+\Delta\boldsymbol{l}_{k}-\boldsymbol{u}_{n}\|}
\end{array}
\end{equation}
For a fixed real-valued vector $\boldsymbol{b}$ and a variable $\boldsymbol{x}\rightarrow\boldsymbol{0}$,
$\|\boldsymbol{b}+\boldsymbol{x}\|$ can be well-approximated by its
first-order Taylor series expansion. This can be written as:

\begin{equation}
\|\boldsymbol{b}+\boldsymbol{x}\|\approx\|\boldsymbol{b}\|+\nabla_{\boldsymbol{x}=\boldsymbol{0}}\|\boldsymbol{b}+\boldsymbol{x}\|^{T}\boldsymbol{x}=\|\boldsymbol{b}\|+\frac{\boldsymbol{b}^{\top}\boldsymbol{x}}{\|\boldsymbol{b}\|}.
\end{equation}
By exploiting this approximation into $\|\hat{\boldsymbol{l}_{k}}+\Delta\boldsymbol{l}_{k}-\boldsymbol{u}_{n}\|$,
we have 
\begin{equation}
\varOmega_{k}^{(n)}(\Delta\boldsymbol{l}_{k})\approx-\frac{\cos\Big(\frac{2\pi}{\lambda}\big(\boldsymbol{q}_{k,n}-\boldsymbol{q}_{k,0}\big)^{T}\Delta\boldsymbol{l}_{k}\Big)}{\|\hat{\boldsymbol{l}_{k}}-\boldsymbol{u}_{n}\|+\boldsymbol{q}_{k,n}^{T}\Delta\boldsymbol{l}_{k}},\label{eq:bound2}
\end{equation}
where $\boldsymbol{q}_{k,n}$ is defined as
\begin{equation}
\boldsymbol{q}_{k,n}=\frac{\hat{\boldsymbol{l}_{k}}-\boldsymbol{u}_{n}}{\|\hat{\boldsymbol{l}_{k}}-\boldsymbol{u}_{n}\|},n=0,\dots,N-1.
\end{equation}
 Although \eqref{eq:bound2} admits a linear form for $\Delta\boldsymbol{l}_{k}$
in both the denominator and the cosine function in the numerator,
it remains mathematically challenging to handle when attempting to
derive an explicit upper bound. Thus, we further consider $\cos x\approx1-\frac{x^{2}}{2}$
for $x\rightarrow0$, we further obtain the approximation of $\varOmega_{k}^{(n)}(\Delta\boldsymbol{l}_{k})$
as 
\begin{equation}
\varOmega_{k}^{(n)}(\Delta\boldsymbol{l}_{k})\approx\frac{\frac{1}{2}\Big(\frac{2\pi}{\lambda}\big(\boldsymbol{q}_{k,n}-\boldsymbol{q}_{k,0}\big)^{T}\Delta\boldsymbol{l}_{k}\Big)^{2}-1}{\|\hat{\boldsymbol{l}_{k}}-\boldsymbol{u}_{n}\|}.
\end{equation}
As a result, $\varOmega_{k}(\Delta\boldsymbol{I}_{k})$ is approximated
as 
\begin{equation}
\varOmega_{k}(\Delta\boldsymbol{I}_{k})\approx\sum_{n=0}^{N-1}\frac{\frac{1}{2}\Big(\frac{2\pi}{\lambda}\big(\boldsymbol{q}_{k,n}-\boldsymbol{q}_{k,0}\big)^{T}\Delta\boldsymbol{l}_{k}\Big)^{2}-1}{\|\hat{\boldsymbol{l}_{k}}-\boldsymbol{u}_{n}\|}.\label{eq:appro}
\end{equation}
Now we are ready to introduce an upper bound for the approximation
in \eqref{eq:appro}.
\begin{prop}
An upper bound for the approximation in \eqref{eq:appro} is given
as 
\begin{equation}
\varepsilon_{k}^{2}\lambda_{\max}(\boldsymbol{Q}_{k})-\sum_{n=0}^{N-1}\frac{2}{\|\hat{\boldsymbol{l}_{k}}-\boldsymbol{u}_{n}\|^{2}},
\end{equation}
where $\lambda_{\max}(\boldsymbol{Q}_{k})$ denotes the maximum eigenvalue
of matrix $\boldsymbol{Q}_{k}\in\mathbb{R}^{2\times2}$ and $\boldsymbol{Q}_{k}$
is defined as 
\begin{equation}
\boldsymbol{Q}_{k}=\frac{4\pi^{2}}{\lambda^{2}}\sum_{n=0}^{N-1}\frac{\Delta\boldsymbol{q}_{k,n}\Delta\boldsymbol{q}_{k,n}^{T}}{\|\hat{\boldsymbol{l}_{k}}-\boldsymbol{u}_{n}\|^{2}},\Delta\boldsymbol{q}_{k,n}=\frac{\hat{\boldsymbol{l}_{k}}-\boldsymbol{u}_{n}}{\|\hat{\boldsymbol{l}_{k}}-\boldsymbol{u}_{n}\|}-\frac{\hat{\boldsymbol{l}_{k}}-\boldsymbol{u}_{0}}{\|\hat{\boldsymbol{l}_{k}}-\boldsymbol{u}_{0}\|}.
\end{equation}
\end{prop}
\begin{IEEEproof}
Please refer to Appendix A.
\end{IEEEproof}
As a result, we obtain an approximated upper bound for the LoS-component
CSI as $\varphi_{k}$ shown at the next page. 
\begin{figure*}
\centering 
\begin{equation}
\varphi_{k}=\frac{\lambda}{4\pi}\sqrt{\varepsilon_{k}^{2}\lambda_{\max}(\boldsymbol{Q}_{k})+\sum_{n=0}^{N-1}\Bigg(\frac{1}{(\|\hat{\boldsymbol{l}_{k}}-\boldsymbol{u}_{n}\|-\varepsilon_{k})^{2}}-\frac{1}{\|\hat{\boldsymbol{l}_{k}}-\boldsymbol{u}_{n}\|^{2}}\Bigg)}\Bigg).\label{eq:pro1}
\end{equation}
\hrule
\end{figure*}

After obtaining the upper bound for the eavesdropping CSI error, the
worst-case eavesdropping SINR constraints in \eqref{eq:C7} and the
worst-case harvested power constraints in \eqref{eq:C8} are respectively
reformulated as 
\begin{eqnarray}
\underset{\|\Delta\boldsymbol{g}_{k}\|\leq\varphi_{k}+\delta_{k}}{\max}\frac{(\hat{\boldsymbol{g}_{k}}+\Delta\boldsymbol{g}_{k})^{H}\hat{\boldsymbol{R}_{0}}(\hat{\boldsymbol{g}_{k}}+\Delta\boldsymbol{g}_{k})}{(\hat{\boldsymbol{g}_{k}}+\Delta\boldsymbol{g}_{k})^{H}\hat{\boldsymbol{R}_{1}}(\hat{\boldsymbol{g}_{k}}+\Delta\boldsymbol{g}_{k})+\xi\sigma_{k}^{2}}\leq\gamma_{R},\nonumber \\
\forall k\in\mathcal{K}_{\mathrm{EAV}},\label{eq:C10}\\
\underset{\|\Delta\boldsymbol{g}_{k}\|\leq\varphi_{k}+\delta_{k}}{\min}\zeta(\hat{\boldsymbol{g}_{k}}+\Delta\boldsymbol{g}_{k})^{H}(\hat{\boldsymbol{R}_{0}}+\hat{\boldsymbol{R}_{1}})\hat{\boldsymbol{g}_{k}}+\Delta\boldsymbol{g}_{k}\geq\xi Q,\nonumber \\
\forall k\in\mathcal{K}_{\mathrm{ER}},\label{eq:C11}
\end{eqnarray}

Next, we introduce the S-procedure to transform the relevant worst-case
constraints into more tractable forms.
\begin{lem}
\label{lem:-S-procedure:-Let-1} S-procedure \cite{ng2014robust,ren2023robust}:
Let $f_{i}(\boldsymbol{e})=\boldsymbol{e}^{H}\boldsymbol{M}_{i}\boldsymbol{e}+2\mathcal{R}e\left\{ \boldsymbol{e}^{H}\boldsymbol{b}_{i}\right\} +n_{i},\boldsymbol{M}_{i}\in\mathbb{S}^{N},\boldsymbol{b}_{i}\in\mathbb{C}^{N\times1},i=1,2$,
and there exists a point $\hat{\boldsymbol{e}}$ such that $f_{1}(\hat{\boldsymbol{e}})<0$.
Then, $f_{1}(\boldsymbol{e})\leq0\Longrightarrow f_{2}(\boldsymbol{e})\leq0$
if and only if there exists $\lambda>0$ such that
\begin{equation}
\lambda\left[\begin{array}{cc}
\boldsymbol{M}_{1} & \boldsymbol{b}_{1}\\
\boldsymbol{b}_{1}^{H} & n_{1}
\end{array}\right]-\left[\begin{array}{cc}
\boldsymbol{M}_{2} & \boldsymbol{b}_{2}\\
\boldsymbol{b}_{2}^{H} & n_{2}
\end{array}\right]\succeq\boldsymbol{0}.
\end{equation}
\end{lem}
Using Lemma \ref{lem:-S-procedure:-Let-1}, the worst-case eavesdropping
SINR constraints in \eqref{eq:C10} are equivalently reformulated
as 
\begin{equation}
\begin{aligned} & \Delta\boldsymbol{g}_{k}^{H}\Delta\boldsymbol{g}_{k}-(\varphi_{k}+\delta_{k})^{2}\leq0\\
\Rightarrow & \Delta\boldsymbol{g}_{k}^{H}\big(\hat{\boldsymbol{R}_{0}}-\gamma_{\mathrm{R}}\hat{\boldsymbol{R}_{1}}\big)\Delta\boldsymbol{g}_{k}+2\mathrm{Re}\Big(\Delta\boldsymbol{g}_{k}^{H}\big(\hat{\boldsymbol{R}_{0}}-\gamma_{\mathrm{R}}\hat{\boldsymbol{R}_{1}}\big)\hat{\boldsymbol{g}_{k}}\Big)\\
 & +\hat{\boldsymbol{g}_{k}}^{H}\big(\hat{\boldsymbol{R}_{0}}-\gamma_{\mathrm{R}}\hat{\boldsymbol{R}_{1}}\big)\hat{\boldsymbol{g}_{k}}-\gamma_{\mathrm{R}}\xi\sigma_{k}^{2}\leq0.
\end{aligned}
\end{equation}

Based on Lemma \ref{lem:-S-procedure:-Let-1}, constraints in \eqref{eq:C10}
are equivalently rewritten as \eqref{eq:LMI constriant} on the top
of next page.
\begin{figure*}
\centering 
\begin{eqnarray}
 & \left[\begin{array}{cc}
\lambda_{k}\boldsymbol{E}-\big(\hat{\boldsymbol{R}_{0}}-\gamma_{\mathrm{R}}\hat{\boldsymbol{R}_{1}}\big) & -\big(\hat{\boldsymbol{R}_{0}}-\gamma_{\mathrm{R}}\hat{\boldsymbol{R}_{1}}\big)\hat{\boldsymbol{g}_{k}}\\
-\hat{\boldsymbol{g}_{k}}^{H}\big(\hat{\boldsymbol{R}_{0}}-\gamma_{\mathrm{R}}\hat{\boldsymbol{R}_{1}}\big) & -\lambda_{k}(\varphi_{k}+\delta_{k})^{2}-\hat{\boldsymbol{g}_{k}}^{H}\big(\hat{\boldsymbol{R}_{0}}-\gamma_{\mathrm{R}}\hat{\boldsymbol{R}_{1}}\big)\hat{\boldsymbol{g}_{k}}+\gamma_{\mathrm{R}}\xi\sigma_{k}^{2}
\end{array}\right]\succeq & \boldsymbol{0},\nonumber \\
 & \lambda_{k}>0,\forall k\in\mathcal{K}_{\mathrm{EAV}}.\label{eq:LMI constriant}
\end{eqnarray}
\hrule
\end{figure*}
 Similarly, the worst-case energy harvesting constraints in \eqref{eq:C11}
are equivalently reformulated as 
\begin{equation}
\begin{aligned} & \Delta\boldsymbol{g}_{k}^{H}\Delta\boldsymbol{g}_{k}-(\varphi_{k}+\delta_{k})^{2}\leq0\\
\Rightarrow & -\Delta\boldsymbol{g}_{k}^{H}\big(\hat{\boldsymbol{R}_{0}}+\hat{\boldsymbol{R}_{1}}\big)\Delta\boldsymbol{g}_{k}-2\mathrm{Re}\Big(\Delta\boldsymbol{g}_{k}^{H}\big(\hat{\boldsymbol{R}_{0}}+\hat{\boldsymbol{R}_{1}}\big)\hat{\boldsymbol{g}_{k}}\Big)\\
 & -\hat{\boldsymbol{g}_{k}}^{H}\big(\hat{\boldsymbol{R}_{0}}+\hat{\boldsymbol{R}_{1}}\big)\hat{\boldsymbol{g}_{k}}+\frac{Q}{\zeta}\xi\leq0.
\end{aligned}
\end{equation}
We can adopt the S-procedure to transform the worst-case harvested
energy constraint in \eqref{eq:C11} as \eqref{eq:LMI constriant-1}
on the next page.
\begin{figure*}
\centering 
\begin{eqnarray}
 & \left[\begin{array}{cc}
\eta_{k}\boldsymbol{E}+\big(\hat{\boldsymbol{R}_{0}}+\hat{\boldsymbol{R}_{1}}\big) & \big(\hat{\boldsymbol{R}_{0}}+\hat{\boldsymbol{R}_{1}}\big)\hat{\boldsymbol{g}_{k}}^ {}\\
\hat{\boldsymbol{g}_{k}}^{H}\big(\hat{\boldsymbol{R}_{0}}+\hat{\boldsymbol{R}_{1}}\big) & -\eta_{k}(\varphi_{k}+\delta_{k})^{2}+\hat{\boldsymbol{g}_{k}}^{H}\big(\hat{\boldsymbol{R}_{0}}+\hat{\boldsymbol{R}_{1}}\big)\hat{\boldsymbol{g}_{k}}-\frac{Q}{\zeta}\xi
\end{array}\right]\succeq & \boldsymbol{0},\nonumber \\
 & \eta_{k}>0,\forall k\in\mathcal{K}_{\mathrm{ER}}.\label{eq:LMI constriant-1}
\end{eqnarray}
\hrule
\end{figure*}
 Therefore, problem (P3) is reformulated as
\begin{eqnarray*}
\hspace{-0.5cm}\textrm{(P4):} & \underset{\hat{\boldsymbol{R}_{0}},\hat{\boldsymbol{R}_{1}},\xi>0,\{\eta_{k}\},\{\lambda_{k}\}}{\max} & \boldsymbol{g}_{0}^{H}\hat{\boldsymbol{R}_{0}}\boldsymbol{g}_{0}\\
 & \textrm{s.t.} & \mathrm{Tr}(\hat{\boldsymbol{R}_{0}}+\hat{\boldsymbol{R}_{1}})\leq\xi P,\\
 &  & \hat{\boldsymbol{R}_{0}}\succeq\boldsymbol{0},\hat{\boldsymbol{R}_{1}}\succeq\boldsymbol{0},\\
 &  & \boldsymbol{g}_{0}^{H}\hat{\boldsymbol{R}_{1}}\boldsymbol{g}_{0}+\xi\sigma_{0}^{2}=1,\\
 &  & \textrm{\eqref{eq:CRB1-1}, \eqref{eq:CRB2-1}, }\eqref{eq:LMI constriant},\textrm{ and }\eqref{eq:LMI constriant-1}.
\end{eqnarray*}
Notice that the CRB constraints in $\eqref{eq:CRB1-1}\textrm{ and }\eqref{eq:CRB2-1}$
and the worst-case constraints in $\eqref{eq:LMI constriant}\textrm{ and }\eqref{eq:LMI constriant-1}$
are now all LMI constraints. It is observed that problem (P4) is a
typical semidefinite programming (SDP) problem that is solvable via
off-the-shelf convex optimization programming solver, such as CVX
\cite{cvx}. 

As a result, by combining the optimal solution of problem (P4) with
a 1D search over $\gamma_{\mathrm{R}}$, we can efficiently derive
a solution to problem (SDR1). 

Notably, the solution to problem (SDR1) may not necessarily satisfy
the rank-one constraint. Conventional rank-one approximation methods
such as Gaussian randomization, may lead to performance degradation.
To address this issue, we present the following theorem to construct
a feasible solution to problem (P1) without any performance degradation
compared to the solution to problem (SDR1).
\begin{prop}
\textup{Let }$\boldsymbol{R}_{0}^{\star}$\textup{ and }$\boldsymbol{R}_{1}^{\star}$\textup{
denote the obtained solution to problem (SDR1), we can always construct
an equivalent solution }$\boldsymbol{R}_{0}^{*}$\textup{ and }$\boldsymbol{R}_{1}^{*}$\textup{
that achieves the same objective value in (P1) and satisfies rank(}$\boldsymbol{R}_{1}^{*}$\textup{)
= 1. 
\begin{equation}
\begin{aligned}\boldsymbol{R}_{0}^{*}=\frac{\boldsymbol{R}_{0}^{\star}\boldsymbol{g}_{0}\boldsymbol{g}_{0}^{H}\boldsymbol{R}_{0}^{\star}}{\boldsymbol{g}_{0}^{H}\boldsymbol{R}_{0}^{\star}\boldsymbol{g}_{0}},\boldsymbol{R}_{1}^{*}=\boldsymbol{R}_{0}^{\star}+\boldsymbol{R}_{1}^{\star}-\boldsymbol{R}_{0}^{*}.\end{aligned}
\end{equation}
}
\end{prop}
\begin{IEEEproof}
Please refer to Appendix B.
\end{IEEEproof}
The complete algorithm to effectively obtain the robust joint beamforming
solution is summarized in Algorithm 1.

\begin{algorithm}

\caption{Robust Joint Beamforming for Solving Problem (P1)}

\begin{itemize}
\item \textbf{Input}:

CU channel vector $\bm{g}_{0}\in\mathbb{C}^{N\times1}$, eavesdropper
locations space$\{\hat{\bm{l}}_{k}\}_{k=1}^{K}$, location uncertainty
$\{\varepsilon_{k}\}_{k=1}^{K}$, NLoS error bounds $\{\delta_{k}\}_{k=1}^{K}$.
\item \textbf{Procedure}:
\begin{itemize}
\item Compute CSI uncertainty bounds $\{\varphi_{k}\}_{k=1}^{K}$ via Proposition
1.
\item Initialize search range $\gamma_{\mathrm{R}}\in[\gamma_{\min},\gamma_{\max}]$.
For each $\gamma_{\mathrm{R}}$, solve problem (P4).
\item Select the optimal solution to obtain $\bm{R}_{0}^{\star}$, $\bm{R}_{1}^{\star}$,
and achievable secrecy rate $C^{\star}$.
\item Apply Proposition 3 to reconstruct rank-one solution to obtain $\boldsymbol{R}_{0}^{*}$
and $\boldsymbol{R}_{1}^{*}$.
\end{itemize}
\item \textbf{Output}: 
\begin{itemize}
\item Information and AN transmit covariance matrix $\boldsymbol{R}_{0}^{*}$
and $\boldsymbol{R}_{1}^{*}$, and achievable secrecy rate $C^{\star}$.
\end{itemize}
\end{itemize}
\end{algorithm}

\subsection{Complexity Analysis}

In this subsection, we analyze the computational complexity of our
proposed robust joint beamforming solution for problem (P1) following
the complexity analysis framework in \cite{6891348}. Given a 1D search
accuracy $\varepsilon_{1\mathrm{D}}>0$, the 1D search iteration complexity
is given as $\mathcal{O}(\frac{1}{\varepsilon_{1\mathrm{D}}})$. Next,
we consider the computational complexity of solving problem (P4).
Specifically, the number of optimization variables is on the order
of $2N^{2}$, and problem (P4) involves $2M$ LMI constraints with
size $2$, 2 LMI constraints with size $N$, $2K$ LMI constraints
with size $N+1$. According to \cite{6891348}, the number of iteration
by the interior-point method is $\mathcal{O}(2N\ln(\frac{1}{\varepsilon}))$,
where $\varepsilon>0$ is the duality gap accuracy. Then, based on
\cite{6891348}, we consider the per-iteration cost of the interior-point
method as $\mathcal{O}\Big(N^{4}M+N^{5}+N^{4}K\Big)$. Finally, we
have the total computational complexity as $\mathcal{O}\Big(N\ln(\frac{1}{\varepsilon})\cdot\frac{1}{\varepsilon_{1\mathrm{D}}}(N^{4}M+N^{5}+N^{4}K)\Big)$.
It is observed that the proposed solutions exhibits polynomial complexity,
that is suitable for practical implementation.

\section{Numerical Results}

In this section, we provide numerical results to validate the effectiveness
of our proposed robust joint secure beamforming design for the near-field
ISCAP system. We assume that the BS is equipped with $N=64$ antennas
and the carrier frequency is set as $28\textrm{ GHz}$ such that $\lambda=0.017\textrm{ m}$.
Considering half-wavelength spacing, we have $d=0.85\textrm{ cm}$
and the Rayleigh distance is around $35\textrm{ m}$ \cite{zhang2024physical}.
To better illustrate the beamforming performance in the angle and
distance domain, we adopt the near-field LoS channel model for the
CU. Specifically, we set the angles of CU and target to be identical,
i.e., $\theta_{s}=\theta_{0}=\frac{\pi}{2}$, the distances of CU
and target is set as $r_{0}=10\textrm{ m}$ and $r_{s}=5\textrm{ m}$,
respectively. $K=3$ ERs are located in $(0\textrm{ m},15\textrm{ m})$,
$(-5\textrm{ m},10\textrm{ m})$, and $(5\textrm{ m},10\textrm{ m})$,
respectively. The uncertain region of each eavesdropper is set as
$\varepsilon_{k}=0.1\textrm{ m}$ and the error bound for NLoS component
is set as $5\%$ of the magnitude of the LoS component. The total
transmit power is set as $P=30\textrm{ dBm}$. Furthermore, the CRB
thresholds for angle and distance are set as $\Gamma_{\theta}=\Gamma_{r}=0.1$
and the harvested power threshold is set as $Q=0.1\textrm{ \ensuremath{\mu}W}$.
The noise power is set as $\sigma_{0}^{2}=\sigma_{k}^{2}=-50\textrm{ dBm}$.

For comparison, we evaluate the following three benchmarks.
\begin{itemize}
\item \textbf{Separate beamforming} \textbf{design}: The sensing/powering/AN
covariance matrix $\boldsymbol{R}_{1}$ is first designed with the
minimum transmit power to satisfy both the CRB constraints for target
sensing and the energy harvesting requirements. Then, with the remaining
power budget, the information beamforming vector $\boldsymbol{w}_{0}$
is optimized to maximize the worst-case secrecy rate. 
\item \textbf{Maximum ratio transmission (MRT)-based benchmark}: The information
beamforming follows the channel direction: $\boldsymbol{R}_{1}=p_{0}\boldsymbol{g}_{0}\boldsymbol{g}_{0}^{H}$,
where $p_{0}$ denotes the allocated power for information transmission.
This approach inherently satisfies the rank-one constraint, thus avoiding
the need for SDR.
\item \textbf{Zero-forcing (ZF)-based benchmark}: The AN covariance matrix
is constrained to lie in the null space of the communication channel,
i.e., $\boldsymbol{g}_{0}^{H}\boldsymbol{R}_{1}=\mathbf{0}$. This
benchmark eliminates interference at the CU, thereby simplifying the
optimization by removing the fractional programming complexity.
\end{itemize}
To begin with, we evaluate the effectiveness of the proposed approximate
upper bound for channel errors. For gaining more insights, we compare
three quantities in the following

\begin{itemize}
\item \textbf{\textcolor{black}{The actual channel errors}} are computed
via Monte Carlo simulation with 10,000 random channel realizations
within the uncertainty region by \eqref{eq:oribound}. Then, we adopt
the maximum channel error as the actual channel error bound.
\item \textbf{\textcolor{black}{The approximate channel errors}} are computed
via Monte Carlo simulation with 10,000 random channel realizations
but exploiting the approximation in \eqref{eq:bound2}. Then, we adopt
the maximum channel error as the approximated channel error.
\item \textbf{\textcolor{black}{Our proposed upper bounds }}\textcolor{black}{are}
computed from Proposition 1 by \eqref{eq:pro1}.
\end{itemize}
In this case, the performance metric is the normalized channel error
bound percentage, defined as 
\[
(\|\boldsymbol{g}_{k}-\hat{\boldsymbol{g}_{k}}\|/\|\hat{\boldsymbol{g}_{k}}\|).
\]

\begin{figure}
\vspace{-0.4cm}\centering\includegraphics[scale=0.38]{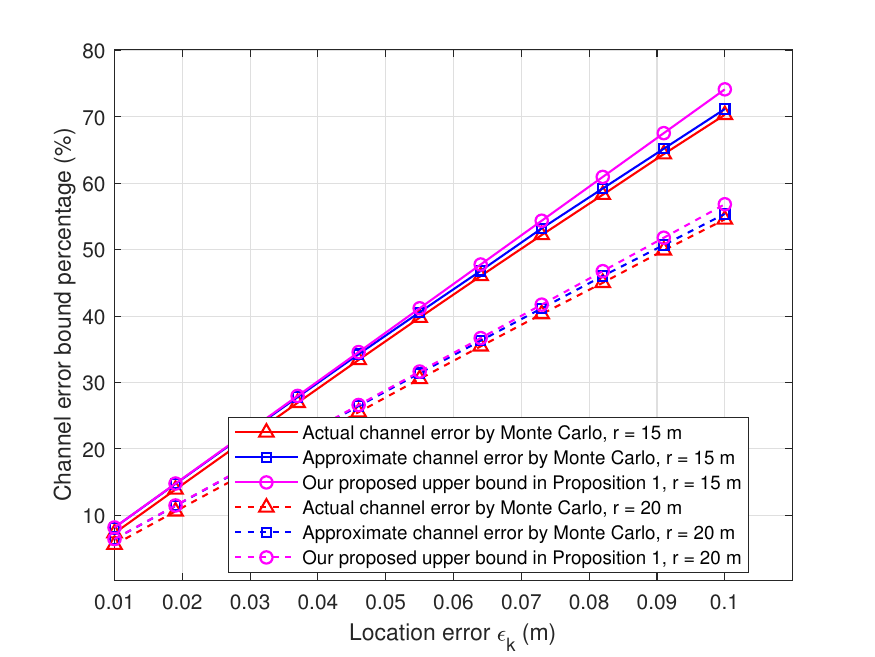}\caption{\label{fig:Channel-error-achieved}Channel error achieved by Monte
Carlo simulation and approximation bound.}
\vspace{-0.4cm}

\end{figure}

Fig. \ref{fig:Channel-error-achieved} demonstrates the comparison
of achieved normalized channel error versus the location error bound
$\varepsilon_{k}$ by Monte Carlo simulations within the uncertainty
region and our approximate error upper bound at $10\textrm{ m}$ and
$15\textrm{ m}$, respectively. It is observed that a lower location
error can naturally reduce the channel error. The results clearly
demonstrate that our proposed upper bound provides an exact performance
guarantee, maintaining the error gap within 10\% of the actual channel
error and consistently below 5\% in topical operational scenarios,
thereby confirming its effectiveness. Furthermore, the approximation
derived in \eqref{eq:bound2} exhibits remarkable accuracy, validating
its effectiveness for practical implementation. Additionally, comparative
analysis across different transmission distances $r$ reveals an important
characteristic that the influence of fixed location estimation error
diminishes progressively as $r$ increases. 

\begin{figure}
\vspace{-0.2cm}\centering\includegraphics[scale=0.38]{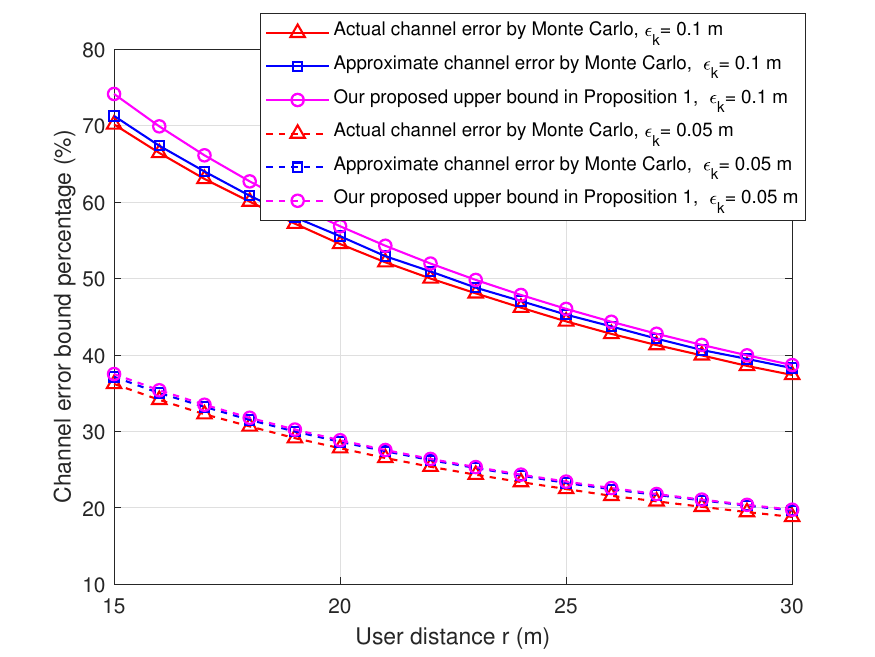}\centering\caption{\label{fig:Channel-error-distance}Channel error achieved by Monte
Carlo simulation and approximation bound.}
\vspace{-0.4cm}
\end{figure}

Fig. \ref{fig:Channel-error-distance} further evaluates the achieved
channel error versus the user distance $r$ for two different location
error values $\varepsilon_{k}$ being $0.05\textrm{ m}$ and $0.1\textrm{ m}$.
First, the proposed upper bound demonstrates tightness across varying
$r$, consistently staying within 5\% of empirical error values. Second,
for a fixed $\varepsilon_{k}$, the channel error decays significantly
with increasing $r$. This trend aligns with intuitive expectations,
as the relative impact of position uncertainty diminishes in larger
distance scenarios. 

\begin{figure}
\vspace{-0.4cm}\centering\includegraphics[scale=0.39]{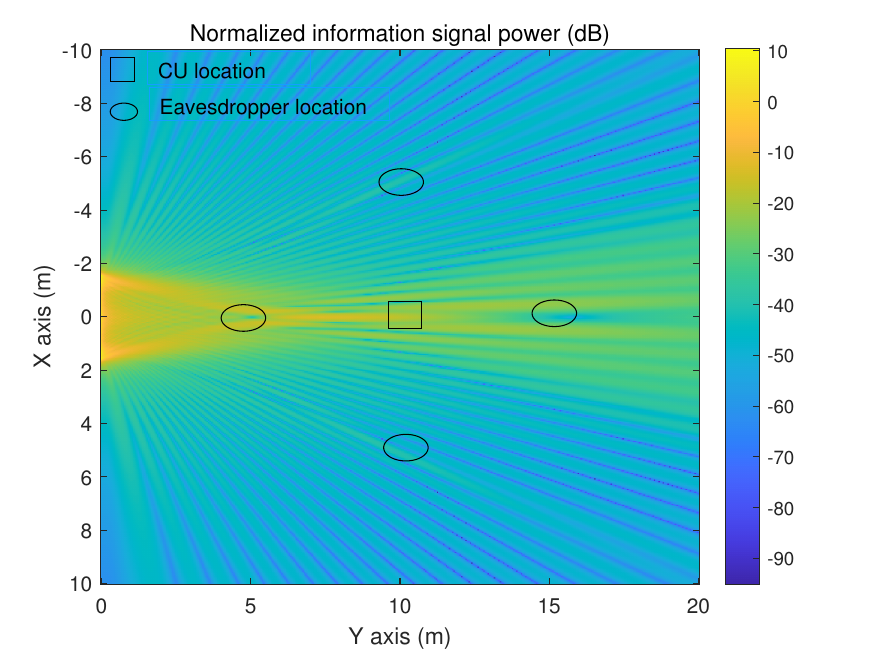}\centering\caption{\label{fig:Normalized-received-communicatio}Normalized received information
signal power.}
\vspace{-0.4cm}
\end{figure}

Next, we demonstrate the spatial power distribution achieved by our
proposed robust joint beamforming scheme. Note that since the sensing
target is located along the same angular direction as the CU, traditional
far-field beamforming cannot ensure secure communications \cite{zhang2024physical}.
However, in our near-field ISCAP system, we leverage near-field transmit
beamforming in the distance domain to enable secure transmissions.
Additionally, considering the critical channel conditions, we set
the location error of the sensing target to $\varepsilon_{k}=0.02\textrm{ m}$.
Fig. \ref{fig:Normalized-received-communicatio} illustrates the normalized
received information signal power distribution in the near-field region,
computed by $\boldsymbol{v}^{H}(\boldsymbol{l})\boldsymbol{R}_{0}^{*}\boldsymbol{v}(\boldsymbol{l})$.
In this figure, the rectangle represents the location of the CU, while
ellipses represent the positions of potential eavesdroppers. It is
observed that near-field beamforming leverages distance-domain spatial
degrees of freedom to achieve pencil-like sharp energy focusing at
the legitimate user's location, ensuring physical layer security when
an eavesdropper shares the same angular direction even is positioned
at a closer distance to the BS. Moreover, by exploiting these distance-domain
degrees of freedom, near-field beamforming can actively generate transmission
nulls to effectively suppress eavesdropping threats. 

\begin{figure}
\centering\includegraphics[scale=0.39]{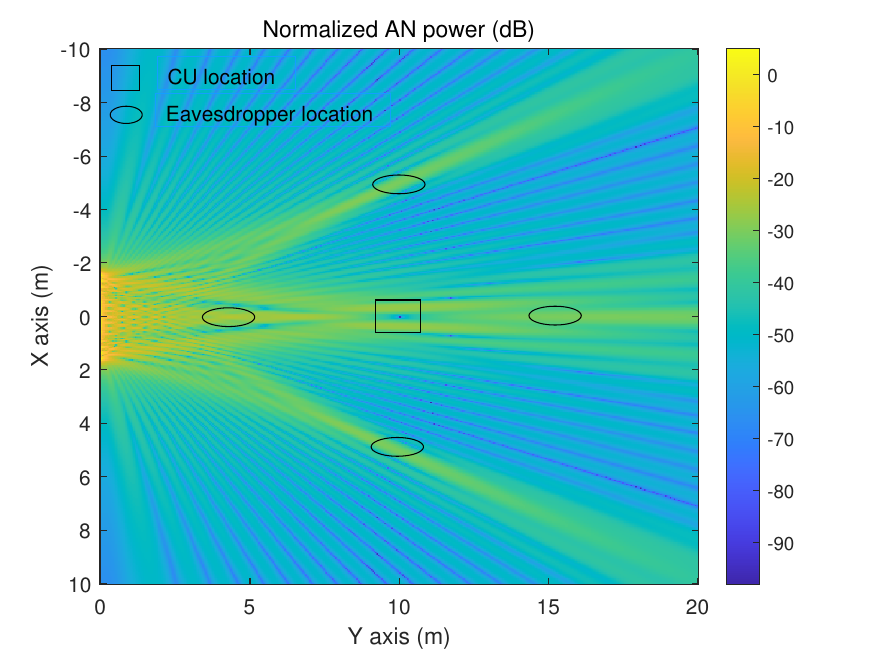}\centering\caption{\label{fig:Normalized-received-AN}Normalized received AN signal power.}
\vspace{-0.4cm}
\end{figure}

Fig. \ref{fig:Normalized-received-AN} illustrates the spatial distribution
of normalized AN power across different locations in the near-field
region, computed by $\boldsymbol{v}^{H}(\boldsymbol{l})\boldsymbol{R}_{1}^{*}\boldsymbol{v}(\boldsymbol{l})$.
First, we observe that near-field beamforming can precisely shape
the AN beam directed toward the ERs or sensing target, enhancing energy
harvesting and sensing performance in the ISCAP system. Simultaneously,
the AN beamforming strategy deliberately forms a spatial null at the
CU's location, effectively eliminating interference.

\begin{figure}
\begin{minipage}[b]{4cm}
    \centering\includegraphics[scale=0.3]{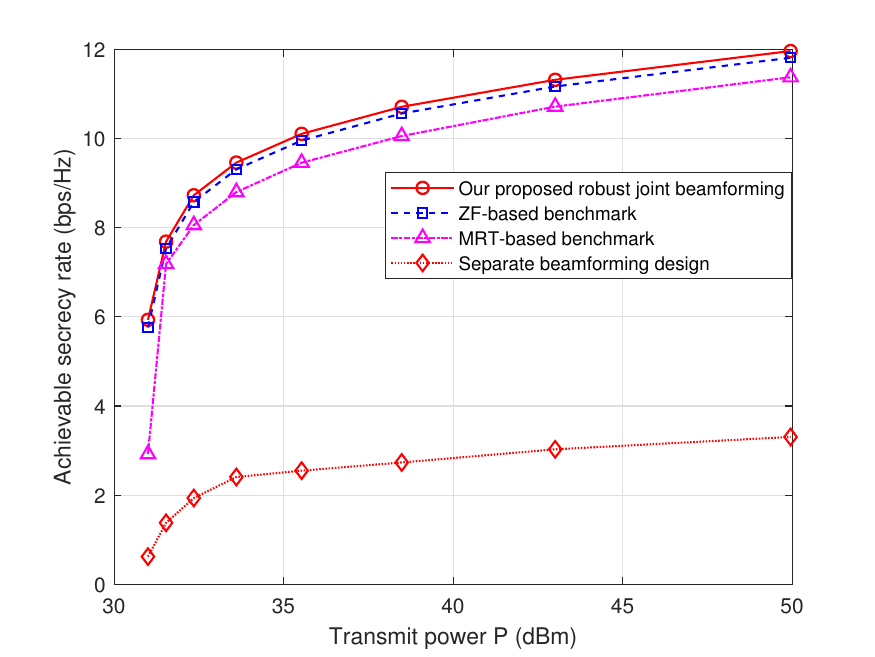}\centering\caption{\label{fig:rate_power}Achievable secrecy rate versus transmit power
$P$.}
\end{minipage}\hfill\begin{minipage}[b]{4cm}
    \centering\includegraphics[scale=0.3]{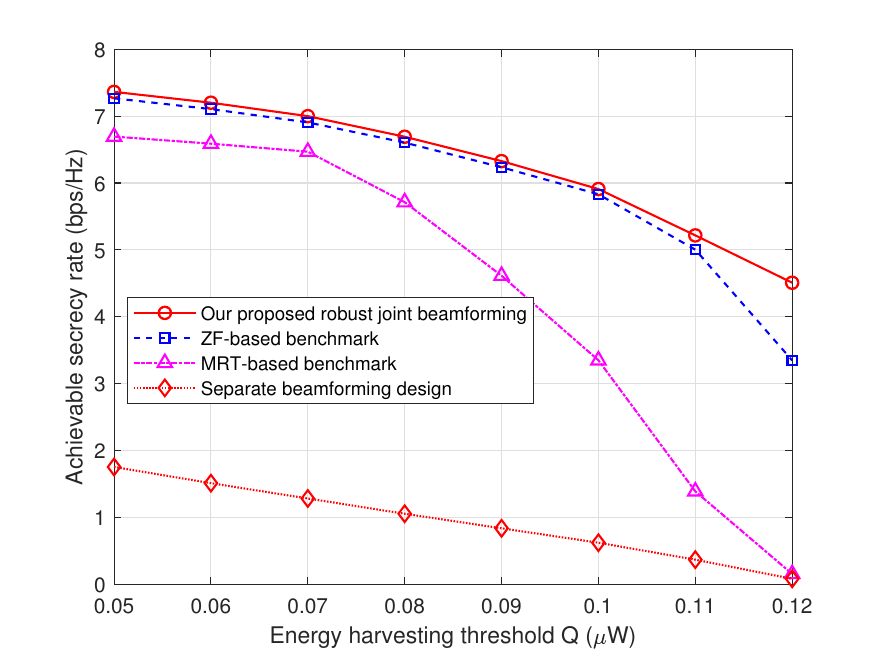}\centering\caption{\label{fig:rate_energy}Achievable secrecy rate versus energy harvesting
threshold $Q$.}
\end{minipage}

\vspace{-0.4cm}
\end{figure}

Fig. \ref{fig:rate_power} shows the achievable secrecy rate versus
transmit power $P$. As the transmit power increases, the achievable
secrecy rate consistently demonstrates a monotonic improvement, with
our proposed robust joint beamforming consistently outperforming all
three benchmarks (separate design, ZF-based, and MRT-based). Notably,
the separate design yields the lowest secrecy rate due to its strictly
suboptimal separate optimization of communication and AN signals,
while the ZF-based benchmark surpasses MRT-based beamforming, leveraging
its inherent interference nulling capabilities and benefiting from
sufficient degrees of freedom ($N=64$ antennas) for effective implementation,
though both remain inferior to our proposed joint design.

Furthermore, Fig. \ref{fig:rate_energy} shows the achievable secrecy
rate versus the energy harvesting threshold $Q$. As the energy harvesting
threshold $Q$ increases, all considered methods experience a decline
in achievable secrecy rate, demonstrating the fundamental non-trivial
trade-off between power transfer and secure communication performance.
Despite this trend, the proposed robust joint beamforming maintains
superior performance across all thresholds, showcasing its efficiency
and robustness. Meanwhile, the ZF-based benchmark delivers moderate
yet stable performance, while the MRT-based method suffers a sharp
degradation, particularly under high thresholds, as the system becomes
interference-limited when nearly all power must satisfy stringent
energy harvesting demands. Under these conditions, the ZF-based benchmark
effectively suppresses interference through null steering, while the
MRT-based benchmark fails to mitigate interference due to its inherent
power-focused beamforming design. The ZF-based benchmark achieves
sufficient DoFs for robust beamforming design, enabled by the massive
spatial-domain resources of ELAA systems. Separate beamforming design
has the lowest secrecy rate and exhibits a noticeable drop as thresholds
increase, which struggles to achieve an effective trade-off between
energy harvesting and secrecy rate.
\begin{figure}
\vspace{-0.4cm}\includegraphics[scale=0.38]{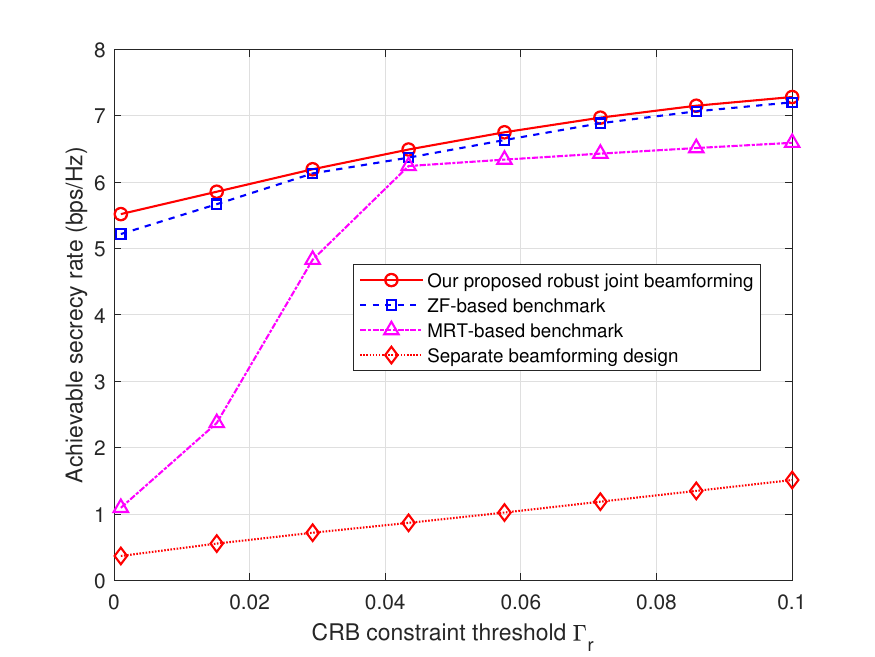}\centering\caption{\label{fig:rate_sensing}Achievable secrecy rate versus CRB constraint
threshold threshold $\Gamma_{r}$.}
\vspace{-0.4cm}
\end{figure}

Besides, Fig. \ref{fig:rate_sensing} shows the achievable secrecy
rate versus CRB constraint threshold $\Gamma_{r}$. It is shown that
as the CRB constraint threshold $\Gamma_{r}$ increases (indicating
looser sensing requirements), all methods exhibit an improvement in
achievable secrecy rate. Similar to Fig. \ref{fig:rate_energy}, the
ZF-based benchmark maintains moderate yet stable performance, whereas
the MRT-based method shows significant degradation under strict CRB
constraints. In such cases, the ZF-based approach successfully mitigates
interference through precise null steering, while the MRT-based benchmark
underperforms due to its fundamental power-focused beamforming architecture. 

\begin{figure}
\begin{minipage}[b]{4cm}
    \includegraphics[scale=0.2]{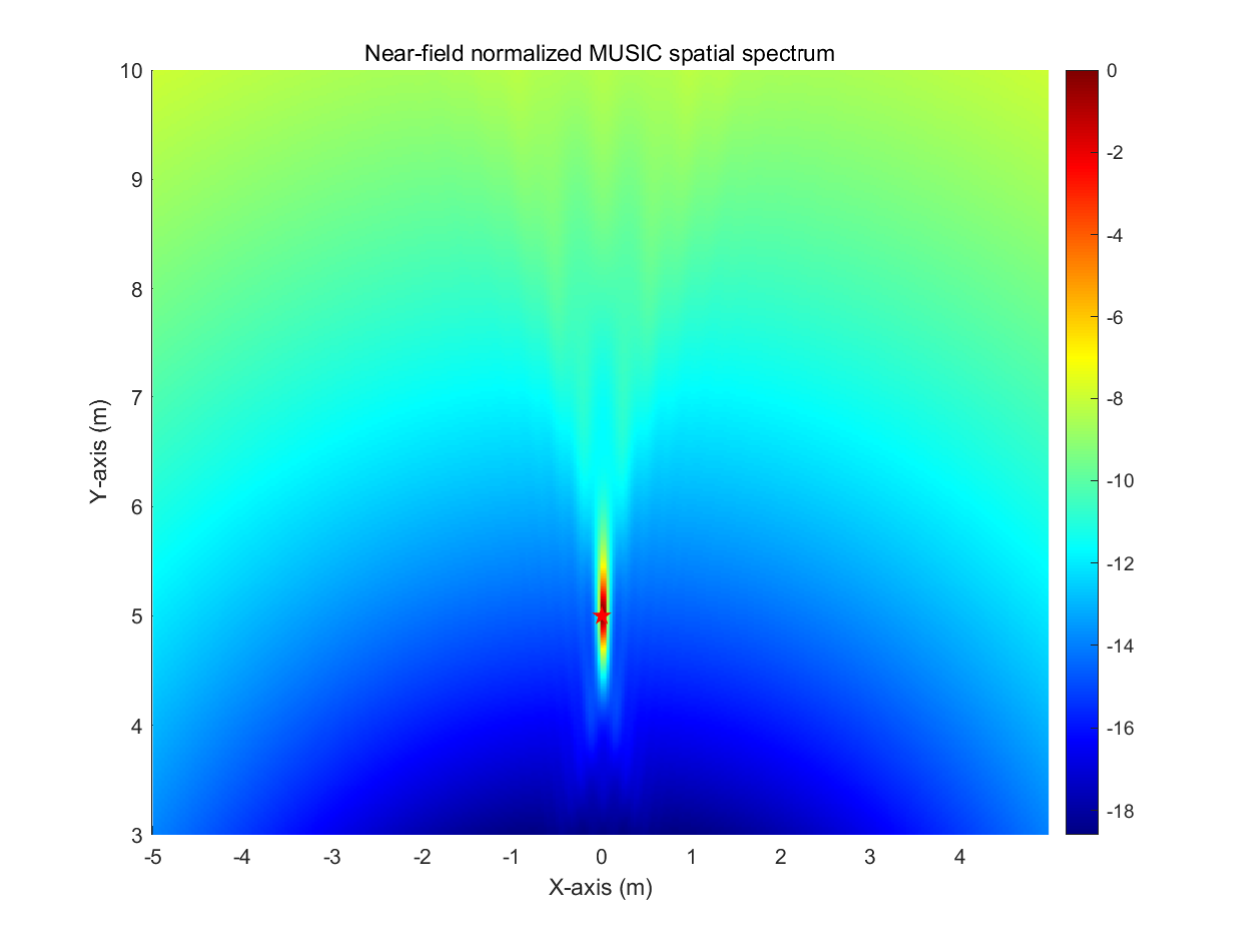}\centering\caption{\label{fig:rMUSIC}Normalized near-field normalized MUSIC spatial
spectrum.}
\end{minipage}\hfill\begin{minipage}[b]{4cm}
    \includegraphics[scale=0.28]{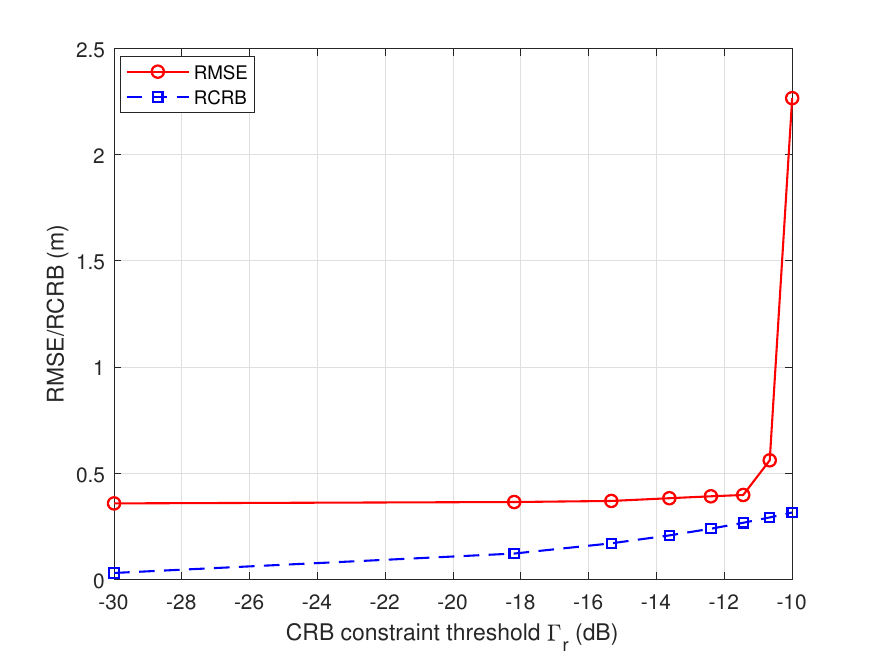}\centering\caption{\label{fig:rate_sensing-1}RCRB/RMSE versus the CRB constraint threshold
$\Gamma_{r}$.}
\end{minipage}\vspace{-0.4cm}
\end{figure}

Furthermore, we conduct target localization via near-field multiple
signal classification (MUSIC) algorithm, detailed in \cite{wang2023near}.
Fig. \ref{fig:rMUSIC} shows the normalized near-filed MUSIC spatial
spectrum. In this figure, the red star marks the location of the sensing
target. The results demonstrate that the near-field MUSIC algorithm
successfully localizes the target by leveraging its high-resolution
capability in the distance domain. Fig. \ref{fig:rate_sensing-1}
shows the localization root mean squared error (RMSE) and root CRB
(RCRB) results corresponding to different CRB constraint threshold.
Each data point represents an average over 1000 random realizations.
The results show that reducing the CRB threshold leads to decreases
in the MUSIC estimation RMSE, confirming the effectiveness of the
sensing design in this ISCAP scenario.

\section{Conclusion}

This paper presented a robust joint beamforming framework for near-field
ELAA systems that maximizes the secrecy rate while guaranteeing worst-case
sensing and energy harvesting performance requirements. By strategically
designing auxiliary beams that serve as energy signals, sensing signals,
and AN, our approach exploits near-field spatial focusing to substantially
degrade the channel quality of potential eavesdroppers. The challenging
non-convex optimization problem was systematically solved through
a combination of SDR, Charnes-Cooper transformation for fractional
programming, and a novel CSI uncertainty characterization method that
decomposes location errors into geometric and NLoS components using
Taylor series approximations. The proposed solution guarantees robustness
through S-procedure-based conversion of worst-case constraints into
tractable LMIs, while maintaining secrecy performance via rigorous
mathematical proofs of SDR tightness. Numerical results demonstrated
significant secrecy rate enhancements while strictly satisfying all
sensing accuracy and power transfer requirements, highlighting the
effectiveness of near-field beamforming in leveraging distance-domain
resolution to simultaneously improve target detection, wireless power
delivery, and secure communications. 

This study opens several promising avenues for future research. First,
advanced beamforming techniques could be developed to handle multiple
distributed sensing targets while maintaining secure communications
and power transfer capabilities. Second, the framework could be extended
to network-level implementations, requiring new network interference
management approaches to preserve security across interconnected nodes.
Third, adaptive algorithms should be investigated to address mobile
scenarios with time-varying channels.

\appendix{}

\section*{Appendix A: Proof of Proposition 1}

We define 
\begin{equation}
\boldsymbol{Q}_{k}=\frac{4\pi^{2}}{\lambda^{2}}\sum_{n=0}^{N-1}\frac{\Delta\boldsymbol{q}_{k,n}\Delta\boldsymbol{q}_{k,n}^{T}}{\|\hat{\boldsymbol{l}_{k}}-\boldsymbol{u}_{n}\|^{2}},\Delta\boldsymbol{q}_{k,n}=\frac{\hat{\boldsymbol{l}_{k}}-\boldsymbol{u}_{n}}{\|\hat{\boldsymbol{l}_{k}}-\boldsymbol{u}_{n}\|}-\frac{\hat{\boldsymbol{l}_{k}}-\boldsymbol{u}_{0}}{\|\hat{\boldsymbol{l}_{k}}-\boldsymbol{u}_{0}\|}
\end{equation}
and obtain the approximation of $\varOmega_{k}(\Delta\boldsymbol{l}_{k})$
as 
\begin{equation}
\varOmega_{k}(\Delta\boldsymbol{l}_{k})\approx\Delta\boldsymbol{l}_{k}^{T}\boldsymbol{Q}_{k}\Delta\boldsymbol{l}_{k}-\sum_{n=0}^{N-1}\frac{2}{\|\hat{\boldsymbol{l}_{k}}-\boldsymbol{u}_{n}\|^{2}}.\label{eq:approximation}
\end{equation}
To determine the upper bound of \eqref{eq:approximation}, we maximize
the expression by solving the following problem: 
\begin{eqnarray}
 & \underset{\Delta\boldsymbol{l}_{k}}{\max} & \Delta\boldsymbol{l}_{k}^{T}\boldsymbol{Q}_{k}\Delta\boldsymbol{l}_{k}\nonumber \\
 & \textrm{s.t.} & \|\Delta\boldsymbol{l}_{k}\|\leq\varepsilon_{k}.\label{eq:Qkproblem}
\end{eqnarray}
 The eigenvalue decomposition of $\boldsymbol{Q}_{k}$ is given as
\[
\boldsymbol{Q}_{k}=\boldsymbol{U}_{k}\boldsymbol{\Lambda}_{k}\boldsymbol{U}_{k}^{T},
\]
where $\boldsymbol{U}_{k}\in\mathbb{C}^{N\times N}$ is an orthogonal
matrix of eigenvectors and $\boldsymbol{\Lambda}_{k}=\textrm{diag}(\lambda_{1},\lambda_{2},\dots,\lambda_{N})\in\mathbb{R}^{N\times N}$
contains the eigenvalues in descending order $\{\lambda_{1}\geq\lambda_{2}\geq\dots,\geq\lambda_{N}\}$.
The optimal value of the quadratic form under the constraint is achieved
when aligns with the eigenvector corresponding to the largest eigenvalue
$\lambda_{1}$. Thus, the maximum value is $\varepsilon_{k}^{2}\lambda_{1}$.
This completes the derivation of the upper bound.

\section*{Appendix B: Proof of Proposition 3}

It is observed that the sensing and energy harvesting performance
metrics depend solely on the summation $\boldsymbol{R}_{0}^{\star}+\boldsymbol{R}_{1}^{\star}$.
Notably, the covariance reconstruction satisfy the equality that $\boldsymbol{R}_{0}^{\star}+\boldsymbol{R}_{1}^{\star}=\boldsymbol{R}_{0}^{*}+\boldsymbol{R}_{1}^{*}$,
which ensures that the sensing accuracy and energy harvesting constraints
are rigorously satisfied.

Next, based on $\boldsymbol{R}_{0}^{\star}\succeq\boldsymbol{0}$,
we have $\boldsymbol{R}_{0}^{\star}=\bar{\boldsymbol{R}_{0}^{\star}}\bar{\boldsymbol{R}_{0}^{\star}}^{H}$.
Based on this, for any $\boldsymbol{v}\in\mathbb{C}^{N\times1}$,
it follows that 
\begin{equation}
\begin{array}[b]{l}
\boldsymbol{v}^{H}(\boldsymbol{R}_{0}^{\star}-\boldsymbol{R}_{0}^{*})\boldsymbol{v}=\boldsymbol{v}^{H}\boldsymbol{R}_{0}^{\star}\boldsymbol{v}-\boldsymbol{v}^{H}\frac{\boldsymbol{R}_{0}^{\star}\boldsymbol{g}_{0}\boldsymbol{g}_{0}^{H}\boldsymbol{R}_{0}^{\star}}{\boldsymbol{g}_{0}^{H}\boldsymbol{R}_{0}^{\star}\boldsymbol{g}_{0}}\boldsymbol{v}\\
=\frac{1}{\boldsymbol{g}_{0}^{H}\boldsymbol{R}_{0}^{\star}\boldsymbol{g}_{0}}(\boldsymbol{v}^{H}\boldsymbol{R}_{0}^{\star}\boldsymbol{v}\boldsymbol{g}_{0}^{H}\boldsymbol{R}_{0}^{\star}\boldsymbol{g}_{0}-\boldsymbol{v}^{H}\boldsymbol{R}_{0}^{\star}\boldsymbol{g}_{0}\boldsymbol{g}_{0}^{H}\boldsymbol{R}_{0}^{\star}\boldsymbol{v})\\
=\frac{1}{\boldsymbol{g}_{0}^{H}\boldsymbol{R}_{0}^{\star}\boldsymbol{g}_{0}}(\|\boldsymbol{a}\|^{2}\|\boldsymbol{b}\|^{2}-\big|\boldsymbol{a}^{H}\boldsymbol{b}\big|^{2})\overset{\textrm{(a)}}{\geq}0,
\end{array}
\end{equation}
where $\boldsymbol{a}=\bar{\boldsymbol{R}_{0}^{\star}}^{H}\boldsymbol{v}\in\mathbb{C}^{\times1},\boldsymbol{b}=\bar{\boldsymbol{R}_{0}^{\star}}^{H}\boldsymbol{g}_{0}\in\mathbb{C}^{N\times1}$,
and inequality (a) holds because of the Cauchy-Schwartz inequality.
Accordingly, we have $\boldsymbol{R}_{0}^{\star}-\boldsymbol{R}_{0}^{*}\succeq\boldsymbol{0}$
and $\boldsymbol{R}_{1}^{*}\succeq\boldsymbol{R}_{1}^{\star}\succeq\boldsymbol{0}$.
Therefore, we have 
\begin{equation}
\frac{\boldsymbol{g}_{0}^{H}\boldsymbol{R}_{0}^{*}\boldsymbol{g}_{0}}{\boldsymbol{g}_{0}^{H}\boldsymbol{R}_{1}^{\star}\boldsymbol{g}_{0}+\sigma_{0}^{2}}=\frac{\boldsymbol{g}_{0}^{H}\boldsymbol{R}_{0}^{*}\boldsymbol{g}_{0}}{\boldsymbol{g}_{0}^{H}\boldsymbol{R}_{1}^{\mathrm{*}}\boldsymbol{g}_{0}+\sigma_{0}^{2}},
\end{equation}

\begin{equation}
\frac{\boldsymbol{g}_{k}^{H}\boldsymbol{R}_{0}^{\star}\boldsymbol{g}_{k}}{\boldsymbol{g}_{k}^{H}\boldsymbol{R}_{1}^{\star}\boldsymbol{g}_{k}+\sigma_{k}^{2}}\leq\frac{\boldsymbol{g}_{k}^{H}\boldsymbol{R}_{0}^{*}\boldsymbol{g}_{k}}{\boldsymbol{g}_{k}^{H}\boldsymbol{R}_{1}^{\mathrm{*}}\boldsymbol{g}_{k}+\sigma_{k}^{2}}.
\end{equation}
Following the proposed reconstruction, the CU SINR remains preserved,
while the eavesdroppers\textquoteright{} SINR is strictly reduced.
Consequently, the secrecy communication performance is guaranteed
to be non-degraded. This completes the proof.

{\footnotesize\bibliographystyle{IEEEtran}
\bibliography{IEEEabrv,IEEEexample,myref}
}{\footnotesize\par}
\end{document}